\documentstyle[draft,psfig]{mn}
\newcommand{\mnref}[1]{\hangindent=0.5in \hangafter=1 #1 \par}
\newenvironment{refs}{\parindent=0pt}{\parindent=1.5em}
\newcommand{\mn}{MNRAS}

\newcommand{\apj}{ApJ}
\newcommand{\apjs}{ApJS}
\newcommand{\aaa}{A\&A}

\newcommand{\Lsolar}{\mbox{\,$\rm L_{\odot}$}}

\title[Near Infrared Spectra of Compact Planetary Nebulae]
{Near Infrared Spectra of Compact Planetary Nebulae}
\author[S.L. Lumsden, P.J. Puxley \& M.G. Hoare]
{S.L. Lumsden$^{1}$, P.J. Puxley$^{2}$ and M.G. Hoare$^{1}$\\
{}$^1$ {\em Department of Physics and Astronomy, University of Leeds, 
Leeds LS2 9JT, UK -- sll@ast.leeds.ac.uk, mgh@ast.leeds.ac.uk}\\
{}$^2$ {\em Gemini Observatory, 670 N. A'ohuku Place, Hilo, Hawaii 96720, USA
-- ppuxley@gemini.edu}\\
}

\begin{document}

\maketitle

\begin{abstract}
This paper continues our study of the behaviour of near infrared helium
recombination lines in planetary nebula.  We find that the 1.7007$\mu$m
4$^{3}$D--3$^{3}$P HeI line is a good measure of the HeI recombination rate,
since it varies smoothly with the effective temperature of the central star.
We were unable to reproduce the observed data using detailed photoionisation
models at both low and high effective temperatures, but plausible explanations
for the difference exist for both.  We therefore conclude that this line
could be used as an indicator of the effective temperature in obscured
nebula.   We also characterised the nature of the molecular hydrogen emission
present in a smaller subset of our sample.  The results are consistent
with previous data indicating that ultraviolet excitation rather than
shocks is the main cause of the molecular hydrogen emission in planetary
nebulae.
\end{abstract}


\section{Introduction}
In a previous paper (Lumsden, Puxley \& Hoare 2001: hereafter Paper 1), we
described how planetary nebulae (PN) could be used to test the use of infrared
helium to hydrogen recombination line ratios as a measure of the effective
temperature of the central star.  Such ratios are often used in infrared
astronomy to constrain the form of the emergent ultraviolet ionising radiation,
and hence to place limits on the  upper end of the stellar initial mass
function in starburst galaxies and HII regions.  PN provide a relatively simple
environment in which to test the behaviour of these lines since there is
generally only one central star present, and they are also visible in the
optical providing a variety of other diagnostic lines to constrain the hardness
of the radiation field.  The temperature of their central stars can therefore
be inferred from standard photoionisation models, unlike, say, the central
stars of obscured compact HII regions (eg Doherty et al.\ 1994).

The fundamental result of Paper 1 was that the bright 2.058$\mu$m
2$^1$P--2$^1$S HeI line, which was previously used to measure the stellar
effective temperature (eg Doyon, Puxley \& Joseph 1992), behaved in a fashion
that was simply too complex to be suitable as a tracer of the size of the
He$^+$ region in the nebula.  Since it is this size relative to the size of the
H$^+$ zone that constraints the incident stellar radiation field, it is clear
that the 2.058$\mu$m line is not a good indicator of the stellar effective
temperature.  Doyon, Puxley \& Joseph had found that the initial mass function
derived from this line in starburst galaxies was anomalous compared to the
results derived from modelling analogous 
bright optical line ratios such as 5007\AA\ [OIII] to H$\beta$.
Clearly such results must now be placed in
doubt.  

However, there is still potential to use HeI/HI line ratios in the way Doyon,
Puxley \& Joseph had intended.  All that is required is a line whose intensity
is dominated by recombination rather than collisional excitation, opacity
driven pumping or fluorescent effects.  We did find that the 2.16465$\mu$m
7$^{3,1}$G--4$^{3,1}$F blend, which is the strongest satellite HeI line to
HI Br$\gamma$, was potentially useful in this fashion.  However, this line will
be completely blended with Br$\gamma$ itself in most galaxies, due to the
smearing caused by the internal velocity dispersion in the system.  The best
other candidate in the near infrared is the 1.7007$\mu$m 4$^{3}$D--3$^{3}$P HeI
line (cf Lumsden \& Puxley 1996, Vanzi \& Rieke 1997).  This paper presents
additional data to Paper 1 allowing us to characterise the behaviour of this
line.

Of course the low resolution spectra we have acquired for this project can be
used in other ways.  The main value, apart from studying the recombination line
spectrum, is in the study of molecular hydrogen emission near PN.  The neutral
and molecular material around PN is largely a tracer of the mass loss that
occurred during the asymptotic giant branch phase.  Molecular hydrogen emission
is more commonly seen in bipolar PN (which form a large fraction of all PN), as
noted by Kastner et al.\ (1996) and Guerrero et al.\ (2000).
Weintraub et al.\ (1998) further suggest that
the onset of molecular hydrogen emission first occurs in proto-planetary
nebulae after the bipolar phase is established but before the central star
reaches a sufficient temperature to ionise the nebula.  Therefore understanding
the molecular hydrogen emission can in principle give important clues as to the
evolution of the molecular envelope during the later proto-planetary phase the
and the early life of the PN itself.  In turn, such evidence may be useful in
constraining the actual mechanisms that give rise to significant asymmetry in
the first place.

There have been many spectroscopic studies of the molecular hydrogen emission
in individual PN in the past.  Detailed examples include Hubble 12, and two of
the objects in our sample, BD+30$^\circ$3639 and CRL~618.  Hubble 12 exhibits a
purely fluorescent spectrum (eg.\ Dinerstein et al.\ 1988, Ramsay et al.\ 1993,
Luhman \& Rieke 1996), as does BD+30$^\circ$3639 (Shupe et al.\ 1998).  CRL~618
appears to show a mixture of fluorescent and collisional excitation (Latter et
al.\ 1992).  However, only Hora, Latter \& Deutsch (1999) have previously
surveyed a large number of PN.  They found that most of the objects they
observed also showed evidence for fluorescent excitation.  Since our sample was
chosen primarily because they were relatively compact on the sky, we have a
sample that is largely unbiased with respect to morphology.  Unfortunately,
because they are compact their actual morphologies are not always well
established.  However, the results from our survey should provide a valuable
extra test of the excitation conditions in the molecular gas.

\section{Observations}
The data presented in this paper were obtained using CGS4, the facility
infrared spectrograph on the United Kingdom Infrared Telescope.  Our sample is
a largely heterogeneous one, composed of those PN that are relatively compact
on the sky and bright enough to allow us to obtain high resolution
spectroscopy, as noted in Paper 1.  The actual sample is given in Table 1.
Other observed parameters, such as electron density etc, are all taken from
Table 1 in Paper 1 and are not reproduced here, with the exception of the
effective temperature, T$_{eff}$.  We have updated the values of the effective
temperature for those objects where T$_{eff}$ is inferred to be greater than
70000K from the presence of significant HeII emission.  We used the strength of
this HeII emission to determine T$_{eff}$ in these cases, as described in
Section 5.2 of Paper 1.  For this paper, we used the observed values of the
2.189$\mu$m HeII 10--7 line rather than the values for the 4686\AA\ HeII line
taken from the literature as in Paper 1.  The revised values are given in Table
4.

We acquired two sets of low resolution ($R\sim400$) spectra.  The first of
these sampled the K waveband between 1.9 and 2.5$\mu$m, using a slit of width
2.4 arcseconds on the sky.  These data were partly included in Paper 1 and a
discussion of the observing details can be found there.  The second set of data
covered the H$+$K waveband between 1.6 and 2.2$\mu$m, providing overlap with
the original K band set.  The H$+$K band data were obtained on the nights of 26
and 27 October 1998 but using a longer focal length camera in the spectrograph.
As a result the slit width is only 1.2 arcseconds on the sky.  In all cases the
slit was oriented north-south on the sky.  We centred the slit on the brightest
infrared portion of the source (which might differ in some cases from the exact
location used in the K band data where we simply centred on the brightest
optical location).

The spectra were obtained in a standard manner, nodding the telescope to keep
the source always on the spectrograph slit.  The beam separation was
approximately 30 arcseconds.  The data were divided by an internal flat field
before all the nod pairs were coadded.  Wavelength calibration was by means of
an internal argon arc lamp.  The arc lamp revealed considerable distortion in
the spectral direction of the raw data.  The image of a strong line on the
array was curved because of distortion within the instrument.  This effect was
calibrated using the arc lines and removed from all the data before further
processing.  The spectra were then extracted, and the negative beam subtracted
from the initial positive one.  This procedure ensures that residual sky
contamination is minimal.

Correction for the atmospheric absorption was achieved using observations of
bright main sequence standard stars.  This process is relatively trivial in the
K band where A type standards can be used after correction for their intrinsic
HI Br$\gamma$ absorption.  In the H band however, the higher Brackett series
lines are all present, and cannot be easily distinguished from absorption due
to our own atmosphere.  We used a variety of spectral types in the range G3--A0
in order to correct for this effect, since their own intrinsic absorption lines
vary between type in a well understood fashion.  The stronger intrinsic
features such as the Brackett lines were fitted using Gaussian or Lorentzian
profiles depending on type.  Weaker atomic and molecular features are seen in
the later spectral types.  These were removed by interpolation on comparison
with the A type standards.  The resulting standard set were cross-checked to
ensure all intrinsic features had been removed.

We used the same stars as crude flux calibrators since we were not seeking
accurate spectrophotometry.  We adopted $V-K$ corrections from Johnson (1966)
and used the $V$ magnitudes from the Bright Star Catalogue (Hoffleit 1982).
This technique is reliable in providing approximate absolute fluxes (accurate
to $\sim$30\%) and very good ($<<5$\%) relative fluxes within each spectra
which are sufficient for our requirements.

All the PN included in this paper were observed at low resolution in the K
band.  We also have high resolution echelle spectroscopy of the kind reported
in Paper 1 for most of our sample.  Those PN without echelle spectroscopy were
not included in Paper 1, since they were not useful for the study of the
2.058$\mu$m 2$^1$P--2$^1$S HeI line.  We did not observe these objects at H$+$K
either.  A few other objects which do have echelle data reported in Paper 1
were not observed at H$+$K because of the hour angle constraints imposed by the
single observing session.

\section{The spectra}
The calibrated H$+$K and K band spectra are presented in Figure 1.  We have not
combined these data since they were obtained with a different aperture.  The
region between 1.8 and 2.0$\mu$m is not shown, as this region is strongly
affected by atmospheric absorption, and little reliable data were obtained.  We
have suppressed the vertical scale in order to reveal the weaker features
present.  We have also not shown the 2--2.2$\mu$m data obtained in 1998 in the
H$+$K data where this duplicates the earlier K band data.  Figure 1(a) shows
the H$+$K band data, Figure 1(b) the K band data and Figure 1(c) the data from
those PN only observed in a single waveband.

We measured line fluxes in these spectra by fitting Gaussian line profiles.
The observed Br$\gamma$ fluxes are given in Table 1 from both the H$+$K and K
band spectra.  We also estimated the average 1$\sigma$ error in the line flux
for all of the lines measured by measuring the residual scatter in the line
subtracted continuum.  These errors are accurate for all regions outside of the
atmospheric absorption bands, and the longest wavelengths of the K band data
where the background noise (which beyond $\sim2.4\mu$m is due to thermal
emission from the telescope) is larger.  These errors are also quoted in Table
1.  The full line fitting results are given in Table 2, after correction for
extinction.  The results from our H$+$K band spectra are given in 2(a), and
from the K band spectra in 2(b).  Values are quoted for all lines which had
more than 4$\sigma$ significance: an indication of possible weaker lines that
may be present is shown by a p.  The data are scaled to the Br$\gamma$ flux in
the same spectrum.  This is typically the brightest line in both spectral
ranges.  The fluxes from the low resolution K band data reported in Paper 1
were remeasured to ensure consistency, and therefore may vary slightly from the
values given there.

We derived extinction measures from the H$+$K band data by comparing the higher
Brackett series lines with Br$\gamma$ where these lines were visible.  We
assumed the extinction law of Landini et al.\ (1984), which varies with
wavelength according to $\tau_\lambda = \tau_0 \lambda^{-1.85}$, and the
theoretical data of Storey \& Hummer (1995).  We did not use the 1.6412$\mu$m
12--4 HI line since it is blended with the 1.644$\mu$m
a$^4$F$_{9/2}$--a$^4$D$_{7/2}$ [FeII] line at the resolution of these data.  We
averaged the results from using the 13--4, 11--4 and 10--4 lines, excluding any
single data point which was clearly discrepant with the other two.  The
resultant values, listed as $\tau_{{\rm Br}\gamma}$, are given in Table 1.
Those values marked with a $\dagger$ were estimated using H$_2$ lines arising
from a common upper level (the 1--0 Q(2) and 1--0 S(0) lines, and the 1--0 Q(3)
and 1--0 S(1) lines).  For most of the PN the results agreed with the Brackett
series lines.  Those noted either did not have the higher Brackett series
lines, or, for the objects showing extended molecular hydrogen emission around
the central HII region, showed significant deviation from the results obtained
from the nebular gas.

One of the objects studied here, BD+30$^\circ$3639, has a dominant Wolf-Rayet
central star which we discuss further in Section 6.  We have excluded the
central portion of this object, labelled BD+30$^\circ$3639 Core in the Tables,
from our analysis.  The WC star has many strong carbon, hydrogen and helium
lines (including the 1.7007$\mu$m 4$^{3}$D--3$^{3}$P HeI line) arising from its
stellar wind (cf the spectra of WC stars in Eenens, Williams and Wade 1991),
rather than from recombination.  However, the region marked in Table 1 as
BD+30$^\circ$3639 Nebula is sufficiently removed from the star that it
represents the true nebular emission.  We uses this region in the analysis of
the helium and hydrogen recombination lines.

\section{The recombination lines}
The spectra shown in Figure 1 are largely dominated by recombination lines of
hydrogen and helium.  In Paper 1 we considered the behaviour of the 2.058$\mu$m
2$^1$P--2$^1$S HeI line at length, as well as the weaker satellite HeI lines
near Br$\gamma$ (in particular the 2.16465$\mu$m 7$^{3,1}$G--4$^{3,1}$F blend)
and the 2.113$\mu$m 4$^{3,1}$S--3$^{3,1}$P HeI blend.  Our intention there was
twofold: to determine how closely each of these lines followed the predictions
of a pure recombination model; and to determine how well each line could be
modelled in a `correct' photoionisation code such as Cloudy (we used version
90.05 for the models in both Paper 1 and this paper -- Ferland 1996).  Both of
these requirements are necessary if we are to use any HeI/HI line ratio as a
measure of the hardness of the stellar spectrum.  

Our main interest in acquiring additional H$+$K band data was to characterise
the behaviour of the 1.7007$\mu$m 4$^{3}$D--3$^{3}$P HeI line.  The observed
ratio of this line with Br$\gamma$ after correction for extinction is given in
Table 4.  Table 4, together with Table 2 from Paper 1, presents the full
results of our observations of near infrared helium-to-hydrogen line ratios in
our PN sample.  In Figure 2(a) we show how the extinction corrected ratio of
the 1.7007$\mu$m 4$^{3}$D--3$^{3}$P HeI line with Br$\gamma$ compares to the
2.16475$\mu$m 7$^{3,1}$G--4$^{3,1}$F HeI to Br$\gamma$ ratio (taken from Paper
1: we scale by Br$\gamma$ since the 2.16475$\mu$m line was measured from our
echelle spectra, and we plot the ratio derived from the echelle spectrum).  The
solid line is the predicted value obtained for recombination processes only
from Benjamin, Skillman \& Smits (1999) and Smits (1996).

By contrast with the relatively clean behaviour of the 1.7007$\mu$m line we
show the behaviour of the 2.113$\mu$m 4$^{3,1}$S--3$^{3,1}$P HeI to Br$\gamma$
ratio in Figure 2(b) (which can be seen as an analogue of the behaviour shown
in Figure 5(b) in Paper 1).  As noted in Paper 1 the 2.113$\mu$m line can be
enhanced by both collisional excitation and opacity in the $2^{3}$S--$n^{3}$P
series.  In this case the solid lines straddle the expected values from
Benjamin, Skillman \& Smits for a low density nebula (ie again the predicted
values assume recombination only).  The dashed line shows the limit when
collisions become important.  Again, theory and observation are in good
agreement, though we cannot say from this plot alone whether the observed
enhancement in those objects that lie to the right of both solid lines is due
to collisions or opacity effects.  In principle we could use the
4$^{3}$P--3$^{3}$D HeI line at 1.95485$\mu$m to test this since it is more
strongly affected by opacity, but the atmospheric transmission near this line
is poor, and deviations of up to a factor of two from the intrinsic values are
possible (cf the values of the Br$\delta$ line in Table 2, which should be
$\simeq$2/3 of the value for Br$\gamma$).  We will therefore not consider this
matter further, other than to say that it is clearly not trivial to model the
behaviour of the 2.113$\mu$m 4$^{3,1}$S--3$^{3,1}$P HeI blend.

Figure 3 shows the behaviour of the 1.7007$\mu$m 4$^{3}$D--3$^{3}$P HeI to HI
Br$\gamma$ ratio as a function of stellar effective temperature.  Also shown
for comparison is the 2.16475$\mu$m 7$^{3,1}$G--4$^{3,1}$F HeI to Br$\gamma$
ratio taken from Paper 1.  Both show a smoothly varying and well behaved trend
as a function of effective temperature.  Also shown are the predictions of our
photoionisation models produced using Cloudy (see Paper 1 for full details).
Two obvious discrepancies appear between model and data for both ratios.  There
is a tendency to underpredict the ratio at both high and low effective
temperature.  At low temperature this discrepancy may be due to the form of the
underlying stellar spectrum assumed.  The central stars of low temperature PN
are not well approximated by black bodies, nor are they well matched by any of
the other possible stellar models in Cloudy at these temperatures.

It is worth noting that the model values shown here differ slightly in their
derivation from those in the equivalent plot for the 2.16475$\mu$m
7$^{3,1}$G--4$^{3,1}$F HeI to Br$\gamma$ ratio in Paper 1 (Figure 9).  The weak
infrared HeI lines are not actually predicted by Cloudy itself.  The optically
thin 4471\AA\ 4$^{3}$D--2$^{3}$P HeI line is predicted by Cloudy however.  We
derived model ratios of the infrared HeI lines with the 4471\AA\
4$^{3}$D--2$^{3}$P HeI line using the data in Benjamin, Skillman \& Smits
(1999) and Smits (1996). The ratio of 1.7007$\mu$m 4$^{3}$D--3$^{3}$P with
4471\AA\ 4$^{3}$D--2$^{3}$P is constant at 0.066 since the two lines arise from
the same level.  The ratio of 2.16475$\mu$m 7$^{3,1}$G--4$^{3,1}$F with
4471\AA\ 4$^{3}$D--2$^{3}$P varies with electron temperature and density.  For
the range of electron temperature and density found in PN this ratio lies
within the bounds 0.026$\pm0.05$.  We also used the Br$\gamma$ value directly
from Cloudy for these plots, whereas in Paper 1 we used the 
predicted H$\beta$ fluxes from Cloudy, and scaled those to predict
Br$\gamma$ fluxes 
using the data of Storey \& Hummer (1995).  It
is notable that the deviation at low effective temperatures seen in Figure 3 is
less with the assumptions made in Paper 1 than with those adopted here.
However, the set we have adopted here are a truer reflection of the predictive
power of Cloudy itself, and should therefore be preferred.

The deviation at high effective temperature is present regardless of which
assumptions we make in deriving the model line ratios.  This implies that we
are overpredicting the fraction of He$^{++}$ in the nebula.  This may indicate
the presence of neutral and partly ionised globules in the nebula, providing an
additional reservoir of neutral and singly ionised helium.  Such globules are a
well known feature of more evolved PN (see, eg, 
Dyson et al.\ 1989 and Meaburn et al.\ 1992).

We also considered the 2.058$\mu$m 2$^1$P--2$^1$S HeI line again, since it
appears in both our H$+$K and K band spectra.  As discussed in detail in Paper
1, the 2.058$\mu$m line lies in a region which suffers from considerable
absorption due to CO$_2$ in our own atmosphere.  These absorption features are
intrinsically very narrow, so that the fraction of line flux absorbed depends
critically on the relative radial velocity of the PN.  If we only correct for
such absorption using the standard star observed at low resolution we will not
correctly account for this fraction, since at low resolution the CO$_2$
absorption features blend together to appear as a broad shallow absorption.  We
can use the observed echelle observations of the 2.058$\mu$m line from Paper 1
to correct for this factor, by comparing the relative transmission in the
observed echelle standard spectra before and after smoothing the data to the
lower resolution achieved with the grating.  We can use the echelle data to
correct our 1998 data as well, since the only relevant factors for any given PN
are the relative radial velocity and the intrinsic line width.  We therefore
followed the same procedure as described in Paper 1 in deriving correction
factors to account for the atmospheric absorption.  These factors are given in
Table 3, along with the corrected 2.058$\mu$m line ratios with Br$\gamma$ (the
observed fluxes are divided by these factors to obtain the correct fluxes).
With the exception of a few objects the results are in very good agreement,
with differences typically less than 10\%, which gives us confidence that our
correction procedure is accurate.  The most discrepant results are for K~4-48,
M~1-12, and SaSt~2-3.  The latter are both very low excitation, and it may
simply be that the data acquired with the narrower slit in 1998 missed some of
the HeI emission present.  We have also adopted an upper limit for M~1-1
as opposed to a weak detection, 
since the H$+$K spectrum shows no clear sign of the 2.058$\mu$m line.
For the purposes of all plots shown, we have
averaged the H$+$K and K data, and used the deviation found between them as a
measure of the true error in our measurement.  The final adopted values
are given in Table 4.  These data supersede the equivalent values given
in Table 2 of Paper 1.

Figure 4 shows the plot of the 2.058$\mu$m 2$^1$P--2$^1$S HeI line ratioed with
HI Br$\gamma$ against stellar effective temperature.  Figure 5 shows the ratio
of the 2.058$\mu$m 2$^1$P--2$^1$S HeI line with the 1.7007$\mu$m
4$^{3}$D--3$^{3}$P HeI line against stellar effective temperature.  The solid
lines shown are the model results presented in Paper 1 and span the likely
parameter space for the conditions present in these PN (see Paper 1 for a
fuller discussion of the models, and the dependence of the 2.058$\mu$m line on
central star luminosity, electron density and intrinsic line width).
Essentially the only difference between the data plotted here and that in Paper
1 is that we have corrected the ratios for extinction, adopted the errors on
the 2.058$\mu$m 2$^1$P--2$^1$S HeI line discussed above, used the 1.7007$\mu$m
4$^{3}$D--3$^{3}$P line as a measure of the intrinsic HeI recombination rate
rather than the 2.16465$\mu$m 7$^{3,1}$G--4$^{3,1}$F blend, and plotted the
temperature data on a log scale to show the full range of the model and
observed behaviour.  We have not changed the models used to ensure consistency
in the comparison.

The results shown in Figure 4 are almost identical to those shown in Figure 10
in Paper 1.  The small shifts in the data that are present actually tend to
make the observations disagree more with the models.  Furthermore, with the
extended temperature scale it can be seen that the observed ratio is larger
than the predicted value for all effective temperatures larger than about
70000K.  Since this parallels the behaviour seen in Figure 3, it is likely that
the cause is the same as alluded to there, namely an underprediction of the
neutral and He$^+$ fractions in these nebula (the former being particularly
important for the 2.058$\mu$m line of course, since it is pumped from resonance
scattering of the 584\AA\ 2$^1$P--1$^1$S line from neutral helium).

This last result is confirmed by Figure 5, which shows that the ratio of two
helium lines does tend to be in better agreement with the models as the
effective temperature increases.  In other regards, Figure 5 is similar to
Figure 11 in Paper 1.  Both Figure 4 and 5 agree with our statement in Paper 1
that the modelling of the 2.058$\mu$m 2$^1$P--2$^1$S HeI in Cloudy is still not
sufficiently accurate for that line to be used as a reliable 
diagnostic.  They also
indicate that our models underpredict the neutral and He$^+$ fractions at much
higher temperatures as noted above.  However, it is clear that whereas the
observed data in Figure 3 follow a relatively well behaved trend as a function
of temperature, the stronger 2.058$\mu$m 2$^1$P--2$^1$S HeI line does not.  We
therefore reiterate our fundamental conclusion of Paper 1, that the 2.058$\mu$m
2$^1$P--2$^1$S HeI line should not generally be used as a measure of stellar
effective temperature.  However, one aspect of Figure 5 is worthy of comment.
It is likely that in specific instances this diagram can be used with Figure 3
to determine if an object has T$_{eff}\gg40000$K.  Examples where this may
be true are where the gas is of sufficiently low density that collisional
excitation of the 2.058$\mu$m line is unimportant (in which case the turbulent
velocity is also likely to be low), or where the actual density and velocity
field of an object are well characterised from other observations.  The
large scatter seen in Figure 5 still indicates that caution should be taken 
in interpreting such results.

Finally, there are several transitions of HeII present in some of our PN.  The
strongest of these are the 10--7 and 12--7 transitions at 2.1891 and
1.6926$\mu$m respectively.  Weaker transitions that can also be seen are the
13--8, 15--8 and 19--8 transitions at 2.348, 2.0379 and 1.7717$\mu$m
respectively.  We have compared the observed fluxes with the expected values
from Storey \& Hummer (1995) and find that they agree well within the errors on
the observational data.  For the range of electron densities and temperatures
seen in these PN there is no particular diagnostic value to these lines, other
than as a measure of the hardness of the stellar radiation field.  Since we
have used the fit to the Stoy temperature from Kaler and Jacoby (1991) to
derive stellar effective temperatures for our sample, and this assumes there is
no O$^{3+}$ or higher ionisation stage of oxygen present, we do not have a
calibrated temperature scale for the HeII emitting PN.  We cannot therefore
test the actual HeII fluxes from the Cloudy models.  Instead in Paper 1 we
inverted this process and used the Cloudy predictions to create a temperature
scale from the observed HeII fluxes for the hotter PNs.  These predictions
must obviously be speculative given our conclusions on the neutral and
singly ionised helium fractions.

\section{Molecular hydrogen emission}
Many of the PN show evidence for some molecular hydrogen emission.  In some
cases, this is limited to a possible weak detection of the 1--0 S(1) line at
2.1218$\mu$m, or the Q-branch beyond 2.4$\mu$m.  These objects are Hu~1-2,
K~3-67, M~1-12, M~1-20 and PC~12.  Of greater interest are the PN that show
sufficient H$_2$ lines to be of diagnostic value.  These are BD+30$^\circ$3639,
CRL~618, K~3-60, K~4-47, K~4-48, M~1-11, M~1-74, M~1-78, M~3-2 and NGC~7027.
Two of these objects, M~1-78 and M~3-2, were previously reported as not showing
molecular hydrogen emission in a survey of PN by Kastner et al.\ (1996).  They
used a 1\% narrow band filter in their survey however, which is only really
suitable for the detection of large equivalent width lines.

The simplest way of characterizing the molecular hydrogen emission is to plot
the observed column density against the energy of the upper level (eg Hasegawa
et al.\ 1987, Ramsay et al.\ 1993).  Figure 6 shows the results for
our data, where we have scaled the column densities by the observed
value for the 1--0 S(1) line.  In this case
\begin{equation}
	\frac{N_{\nu J}}{N_{13}} = 
	\frac{I_{\nu\nu',JJ'}}{I_{10,31}} 
	\frac{A_{10,31}}{A_{\nu\nu',JJ'}}
	\frac{\lambda_{\nu\nu',JJ'}}{\lambda_{10,31}},
\end{equation}
where $N_{\nu J}$ is the column density in the upper level,
$I_{\nu\nu',JJ'}$ is the intensity of the $\nu-\nu'$, J--J$'$
transition, and $A_{\nu\nu',JJ'}$ and $\lambda_{\nu\nu',JJ'}$
are the transition probability and wavelength respectively associated
with that transition.  We used the transition probabilities 
from Turner, Kirby-Docken \& Dalgarno (1977), and took the 
intrinsic wavelengths from Black \& van Dishoeck (1987).

Assuming thermal equilibrium, the column densities are related by
\begin{equation}
	\frac{N_{\nu J}}{N_{\nu' J'}}
	= \frac{g_J}{g_{J'}} \exp \left( 
	-\frac{E(\nu,J) - E(\nu',J')}{kT}\right),
\end{equation}
where $E(\nu,J)$ is the energy of the upper level, and $g_J = g_S(2J+1)$ is the
statistical weight of the upper level.  We used the energy levels tabulated by
Dabrowski (1984).  The weights for the different spin states are $g_S = 1$ for
para H$_2$ (even $J$), and 3 for ortho $H_2$ (odd $J$), though the observed
ortho/para ratio may deviate from this (eg Sternberg \& Neufeld 1999).

Clearly, from equation 2, plotting the logarithm of the ratio of the column
densities divided by the appropriate statistical weights gives the rotational
temperature from the slope of the data.  In practice, although we plot all the
data, we generally only use those lines with wavelengths between 1.6 and
1.75$\mu$m, and 2.0 and 2.4$\mu$m, in deriving the rotational temperature.
These should have the most reliable fluxes.  At the resolution of our data the
many Q-branch lines beyond 2.4$\mu$m are blended together, and with the HI
Pfund series lines. Several also lie near strong atmospheric absorption
features.  The higher 1--0 S($n$) lines in the H band tend to be blended with
atomic features in most of our spectra.  Clearly in such cases the average
errors tabulated in Table 1 are likely to be an underestimate.  Therefore, we
assume for the purpose of Figure 6 that the measurement error is 50\% where the
lines are heavily blended with other species or lie outside the `good'
atmospheric bands.  This does not affect any of our results, since we do not
use these points in any actual analysis. The rotational excitation temperatures
derived from Figure 6 are given in Table 5.

It is also instructive to ratio the observed data with a suitable
model.  The right hand panels in Figure 6 show the results of scaling
all the column densities with the values expected from a model
with low density and pure fluorescent emission, taken from
model 14 of Black \& van Dishoeck (1987).  Although more accurate models
have been calculated for the conditions actually expected in
a photo-dissociation region (see, eg, Draine \& Bertoldi 1996, and
references therein), the basic nature of the Black \& van Dishoeck
model actually makes it easier to distinguish differences between
the observed data.   Clearly, any PN in which the molecular hydrogen emission
arises from fluorescence in a low density environment should be a
reasonable match to the Black \& van Dishoeck model, in which case the 
data will scatter about a horizontal line in the right hand panels.

The molecular hydrogen emission region in BD+30$^\circ$3639 is an example in
which this is approximately true, as are M~1-11 and M~1-74.  This result for
BD+30$^\circ$3639 differs from that found from studies of the bright H$_2$
emitting lobes that lie east/west of the main nebula (the region we describe as
the H$_2$ region in BD+30$^\circ$3639 lies north and south of the main nebula).
Shupe et al.\ (1998) found that their data required an enhancement over the
model in the v=1--0 lines consistent with emission from high density gas.  In
this case the v=1 levels can become thermalised.  The emission from all the
regions of M~1-78 is similar to the behaviour they see.  By contrast the
reflection nebula in CRL~618 shows behaviour consistent with all the detected
lines arising from shock excitation (which is why they align approximately
along a common slope in the left hand panel).  However, both Hora, Latter \&
Deutsch (1999) and Latter et al.\ (1992) detected weak lines arising from
higher levels in the J band, indicating that there must also be some UV
excitation from the central star.  Of the other PN we observed, several do not
have enough diagnostic lines to adequately distinguish shock excitation from
high density UV excitation.  K~4-47 is probably shock excited on the basis of
the analogy with CRL~618.  K~4-48 and M~3-2 could be explained by either model.

Some of the PN may also show evidence for emission from the unidentified lines
discussed by Geballe, Burton and Isaacman (1991).  These coincide in wavelength
with the 3--2 S(2) and 3--2 S(3) lines of H$_2$ at the resolution of our data.
K~3-60 is an example in which this is clearly the case (in other respects the
data for this object are reasonably consistent with a fluorescent excitation
mechanism).  M~1-11 may also show evidence for these lines, as does NGC~7027
(which again in other respects is reasonably consistent with a fluorescent
model).

The other distinguishing factor in the H$_2$ excitation is the observed
ortho/para ratio.  As discussed at length by Sternberg \& Neufeld (1999) this
can take values other than the intrinsic one of 3 if the excitation is due to
UV fluorescence, since the transitions are pumped by optically thick UV
transitions in which the optical depth depends on whether the molecule is in an
ortho or para state.  We followed the procedure outlined in Ramsay et al.\
(1993) in determining the observed ratios.  These are given in Table 5.  We
were able to determine ratios from the v=1--0 levels for all of the objects
discussed above.  Only a few of the PN had sufficient lines to determine
ortho/para ratios from the v=2--1 transitions (the main problem is that the
2--1 S(2) line at 2.15421$\mu$m is blended with other features in PN where the
spectrum is dominated by the ionised gas).  For these cases we list two values
where the data warrants it.  The first is derived from the comparison of the
2--1 S(3) and 2--1 S(2) lines, the second from the 2--1 S(1) and 2--1 S(2)
lines.  Where only one value is given it was derived from the 2--1 S(3) and
2--1 S(2) lines.

The first point to note from our data is that the lowest values of the
ortho/para ratio from the v=1--0 lines all arise from the PN showing relatively
pure fluorescent H$_2$ emission (M~1-74, M~1-11 and BD+30$^\circ$3639)
consistent with the analysis of Sternberg \& Neufeld.  Several of the PN have
ratios consistent with the intrinsic value (indicating shocks or thermalisation
of the v=1 states) consistent with our analysis above.  BD+30$^\circ$3639 also
shows the lowest derived ratio from the v=2--1 lines.  For the other objects
listed in Table 5, both CRL~618 and K~4-47 have ortho/para ratios that are
approximately consistent from both the v=1--0 and 2--1 lines.  This strengthens
the conclusion that the H$_2$ emission in K~4-47 is shock excited, akin to
CRL~618.  The different regions of M~1-78 are perhaps the most interesting
though.  Although they have derived ortho/para ratios from the v=1--0 lines
consistent with the intrinsic value, the ratios derived from the v=2--1 lines
are much lower.  As noted by Sternberg \& Neufeld this is actually what you
would expect if the v=1 levels are thermalised.

Overall our results agree well with data obtained for the same PN by Hora,
Latter \& Deutsch (1999).  We find, as they did, that the majority of PN show
molecular hydrogen emission caused by fluorescent UV excitation by the central
star.

\section{Other features}
There are a variety of other features in some of the spectra which have
no particular diagnostic value in the present data, but are worth mentioning
briefly.  First, we note the presence of many permitted carbon lines in the
stellar spectrum in the core of BD+30$^\circ$3639.  We have compared our
spectrum with those given in Eenens, Williams and Wade (1991).  The best
diagnostic of which type of WC star is present is probably given by the CIV
multiplet near 2.08$\mu$m.  The fact that this multiplet is resolved into its
components is clear evidence in favour of the star having a WC9 spectrum.  This
is identical to the optical classification (eg.\ Smith \& Aller, 1971).

Hu 1-2 shows evidence for emission from the [SiVI] line at 1.9634$\mu$m.  This
is in keeping with its very high excitation.  As many of the other high
excitation PN show molecular hydrogen emission, it is not possible to say if
they also may show the [SiVI] line, since it will be largely blended with the
1--0 S(3) H$_2$ line.

The other relatively common features present are the unidentified lines
discussed by Geballe, Burton and Isaacman (1991).  They can be easily spotted
by comparison with the strength of the 1--0 S(1) H$_2$ line strength as seen in
the previous section.  Objects which show only the emission from the
unidentified lines and no or very weak H$_2$ include K 3-62, M 1-4, M 1-6, M
1-9, M 1-12, M 1-14, Me 2-1 and Vy 1-1.  Geballe et al.\ claimed that the
parent ion responsible for these features corresponds to an excitation level
between 30 and 60eV, but the higher end of this range is inconsistent with the
observed presence of these lines in objects with effective temperatures
$<40000$K.  Dinerstein (2001) has recently claimed an identification for these
feature as [KrIII] and [SeIV] for the 2.199 and 2.287$\mu$m lines respectively.
The abundance of these elements is likely to be higher in PN than in classical
HII regions, since the material could be expelled from the progenitor star.
This identification does agree with the lower effective temperature of a PN
such as M~1-9.  Furthermore, in M~1-12, with an effective temperature of only
26000K, we see only the 2.199$\mu$m feature.  The [KrIII] line arises from a
state with ionisation potential similar to that of He$^+$, which we do see in
this object.  Therefore the observed data are consistent with the
identifications proposed by Dinerstein.  However, as discussed in Lumsden \&
Puxley (1996), we have also be seen these features in at least one classical
HII regions.  The abundance of both krypton and selenium in the interstellar
medium is sufficiently low that this remains a puzzle.  It would be worth
testing whether these lines can be seen in other HII regions when spectra of
sufficiently high signal-to-noise are acquired.

The only other lines clearly present in any of our targets are due to [FeII]
and [FeIII].  Unfortunately, with the exception of CRL 618 the detections are
either too weak, too blended or lacking sufficient lines to allow the data
to be used in determining the physical properties of the nebula.  The H$+$K
band [FeII] lines in CRL 618 are consistent with an electron density larger
than 10000cm$^{-3}$.  This agrees well with the values found by Kelly, Latter
and Rieke (1992) from shorter wavelength data of the reflection nebulosity.

\section{Conclusions}
We have presented new H$+$K band data for the sample of PN we discussed in
Paper 1.  From these data we have been able to show that the 1.7007$\mu$m
4$^{3}$D--3$^{3}$P HeI line is dominated by recombination, and varies smoothly
as a function of the stellar effective temperature of the ionising star.  This
implies that it is a suitable candidate for use in constraining the stellar
initial mass function in the spirit of Doyon, Puxley and Joseph (1992).
Clearly our PN data do not provide an accurate calibration of the expected
4$^{3}$D--3$^{3}$P HeI to HI Br$\gamma$ ratio in star forming regions since the
emergent stellar spectrum is rather different in OB stars and in planetary
nebulae central stars.  The final paper in this sequence will therefore
consider the behaviour of this line in a similar sample of compact HII regions
to those studied by Doherty et al.\ (1994).  Although these data will be
constrained to a far smaller range in effective temperature, they should enable
us to calibrate how this ratio varies in real star forming regions.

We were also able to study the molecular hydrogen emission in a subset 
of our data.  Of the 26 objects we observed, 15 showed some emission
due to molecular hydrogen.  Ten of those had enough detectable H$_2$ lines
to enable us to characterise the emission.  The majority of those showed
evidence for UV excitation as the cause of the emission, in agreement
with the conclusions of Hora, Latter \& Deutsch (1999).

\section*{Acknowledgments}
We would like to thank Derck Smits for providing the machine readable version
of his HeI predictions from his 1996 paper, and the referee, Rene Doyon,
for his useful comments.  SLL acknowledges support from
PPARC through the award of an Advanced Research Fellowship.  SLL also thanks
the Access to Major Research Facilities Program, administered by the Australian
Nuclear Science and Technology Organisation on behalf of the Australian
Government, for travel support for the observations reported here.  The United
Kingdom Infrared Telescope is operated by the Joint Astronomy Centre on behalf
of PPARC.

\section*{References}
\begin{refs}
\mnref{Benjamin, R.A., Skillman, E.D., Smits, D.P., 1999, \apj, 514, 307}
\mnref{Black, J.H., van Dishoeck, E.F., 1987, \apj, 322, 412} 
\mnref{Dabrowski, I., 1984, Canadian J Phys, 62, 1639}
\mnref{Dinerstein, H.L., Lester, D.F., Carr, J.S., Harvey, P.M., 
	1988, \apj, 327, L27} 
\mnref{Dinerstein, H.L., 2001, \apj, in press}
\mnref{Doherty, R.M., Puxley, P.J., Doyon, R.,  Brand, P.W.J.L., 1994,
	\mn, 268, 821}
\mnref{Doyon, R., Puxley, P.J., Joseph, R.D., 1992, \apj, 397, 117}
\mnref{Dyson, J.E., Hartquist, T.W., Pettini, M., Smith, L.J., 
	1989, \mn, 241, 625} 
\mnref{Draine, B.T., Bertoldi, F., 1996, \apj, 468, 269} 
\mnref{Eenens, P.R.J., Williams, P.M., Wade, R., 1991, \mn, 252, 300} 
\mnref{Ferland, G.J., 1996, Hazy, a Brief Introduction to Cloudy, 
	University of Kentucky Department of Physics and Astronomy 
	Internal Report.}
\mnref{Geballe, T.R., Burton, M.G., Isaacman, R., 1991, \mn, 253, 75} 
\mnref{Guerrero, M.A., Villaver, E., Manchado, A., Garcia-Lario, P.,
	Prada, F., 2000, \apjs, 127, 125} 
\mnref{Hasegawa, T., Gatley, I., Garden, R.P., Brand, P.W.J.L., 
	Ohishi, M., Hayashi, M., Kaifu, N., 1987, \apj, 318, L77} 
\mnref{Hoffleit, D., 1982, The Bright Star Catalogue (Fourth Ed.), 
   Yale University Observatory.}
\mnref{Hora, J.L., Latter, W.B., Deutsch, L.K., 1999, \apjs, 124, 195} 
\mnref{Johnson, H.L., 1966, ARA\&A, 4, 193}
\mnref{Kaler, J.B., Jacoby, G.H., 1991, \apj, 372, 215}
\mnref{Kastner, J.H., Weintraub, D.A., Gatley, I., Merrill, K.M., 
	Probst, R.G., 1996, \apj, 462, 777} 
\mnref{Kelly, D.M., Latter, W.B., Rieke, G.H., 1992, \apj, 395, 174}
\mnref{Landini, M., Natta, A., Oliva, E., Salinari, P., Moorwood, A.F.M., 
      1984, \aaa, 134, 284}
\mnref{Latter, W.B., Maloney, P.R., Kelly, D.M., Black, J.H., 
	Rieke, G.H., Rieke, M.J., 1992, \apj, 389, 347} 
\mnref{Luhman, K.L., Rieke, G.H., 1996, \apj, 461, 298} 
\mnref{Lumsden, S.L., Puxley, P.J., 1996, \mn, 281, 493}
\mnref{Lumsden, S.L., Puxley, P.J., Hoare, M.G., 2001, \mn, 320, 83}
\mnref{Meaburn, J., Walsh, J.R., Clegg, R.E.S., Walton, N.A., 
	Taylor, D., Berry, D.S., 1992, \mn, 255, 177} 
\mnref{Ramsay, S.K., Chrysostomou, A., Geballe, T.R., Brand, P.W.J.L.
	Mountain, M., 1993, \mn, 263, 695}
\mnref{Shupe, D.L., Larkin, J.E., Knop, R.A., Armus, L., Matthews, K.,
	 Soifer, B.T., 1998, \apj, 498, 267} 
\mnref{Smith, L.F., Aller, L.H., 1971, \apj, 164, 275} 
\mnref{Smits, D.P., 1996, \mn, 278, 683}
\mnref{Sternberg, A., Neufeld, D.A., 1999, \apj, 516, 371} 
\mnref{Storey, P.J., Hummer, D.G., 1995, \mn, 272, 41}
\mnref{Turner, J., Kirby-Docken, K., Dalgarno, A., 1977, \apjs, 35, 281} 
\mnref{Vanzi, L., Rieke, G.H., 1997, \apj, 479, 694}
\mnref{Weintraub, D.A., Huard, T., Kastner, J.H., Gatley, I., 1998, \apj, 
	509, 728}

\end{refs}

\onecolumn

\tabcolsep=5.5pt
\begin{tabular}{lcrrrrr}
Name & \multicolumn{1}{c}{Extracted Region} & \multicolumn{2}{c}{F(Br$\gamma$)}
 & \multicolumn{2}{c}{1$\sigma$ Error} & \multicolumn{1}{c}{$\tau$(Br$\gamma$)}\\
  &  \multicolumn{1}{c}{(arcsec)} & \multicolumn{2}{c}{ ($\times10^{-18}$Wm$^{-2}$) } 
 & \multicolumn{2}{c}{ ($\times10^{-19}$Wm$^{-2}$) }  \\
  &    & H$+$K & K & H$+$K & K \\
\multicolumn{7}{c}{Compact, or Homogeneous Objects}\\
Hu 1-2 & 18.0  & 66.5 & 205 & 4.8 &   24 & $0.00\pm0.03\phantom{^\dagger}$ \\
K 3-60 &  7.2 & 131 & 183 & 5.4 &   19 & $0.32\pm0.05\phantom{^\dagger}$   \\
K 3-62 &  7.8 & 305 & 706 & 4.9 &   10 & $0.44\pm0.02\phantom{^\dagger}$\\
K 3-66 &  7.2 & 99.7 & 113 & 3.0 &   5 & $0.13\pm0.07\phantom{^\dagger}$\\
K 3-67 &  7.2 & 184 & 182 & 4.3 &   5 & $0.21\pm0.05\phantom{^\dagger}$\\
K 4-48 &  7.2 & 30.8 & 72.7 & 2.0 &   5 & $0.35\pm0.08\phantom{^\dagger}$\\
M 1-1 &  12.6 &  16.1 & 24.1  & 4.0 &  5 & $0.00\pm0.08\phantom{^\dagger}$\\
M 1-4 &  10.8 & 198 & 266  & 3.0 &   5 & $0.38\pm0.05\phantom{^\dagger}$\\
M 1-6 &   9.6 & 310 & 554  & 4.8 &   10 & $0.34\pm0.06\phantom{^\dagger}$\\
M 1-9 &   8.4 & 153 & 211  & 4.8 &   10 & $0.13\pm0.02\phantom{^\dagger}$\\
M 1-11 &  11.4 & 914 & 1130 & 24 &   14 & $0.16\pm0.02\phantom{^\dagger}$\\
M 1-12 &  9.0 & 301 & 381 & 9.1 &   10 & $0.13\pm0.02\phantom{^\dagger}$\\
M 1-14 &  9.0 & 194 & 693 & 4.8 &   24 & $0.13\pm0.04\phantom{^\dagger}$\\
M 1-20 &  7.2 & 286 & 282 & 4.8 &   10 & $0.11\pm0.02\phantom{^\dagger}$\\
M 1-74 &  6.6 & 287 & 329 & 4.8 &   10 & $0.13\pm0.05\phantom{^\dagger}$\\
NGC 7027 & 16.8  & 4440 & 10600 & 140 &   190 & $0.10\pm0.08\phantom{^\dagger}$\\
SaSt 2-3  &7.2 & 25.8 & 38 & 3.3 &    5 & $0.18\pm0.08\phantom{^\dagger}$\\
Vy 1-1 & 9.0 & 47.9 & 110 & 2.9  &  14 & $0.18\pm0.04\phantom{^\dagger}$\\
\multicolumn{7}{c}{Extended Objects }\\
M 1-78 Nebula & 12.6 & 1430 & 910 & 9.5  &  29 & $0.72\pm0.05\phantom{^\dagger}$\\
M 1-78 H$_2$~North & 12.0 & 11.1 & 100 & 14 &   12 & $1.90\pm0.40^\dagger$ \\
M 1-78 H$_2$~South & 6.6  & 16.3 & 119 & 4.8 &   10 &
 $0.72\pm0.05\phantom{^\dagger}$ \\
BD+30$^\circ$3639 Core & 6.0 & 1780 & 3770 & 14 &   48 & $0.10\pm0.02\phantom{^\dagger}$ \\ 
BD+30$^\circ$3639 Nebula  & 3.6+4.2& 357 & 1300 & 14 &   48 &
 $0.10\pm0.02\phantom{^\dagger}$ \\
BD+30$^\circ$3639 H$_2$ zone  & 7.8+10.2& 10.8 & 175 & 10 &    20 &
 $0.10\pm0.02\phantom{^\dagger}$ \\
\multicolumn{7}{c}{Objects lacking H$+$K or K data}\\
CRL 618     & 7.2 & 263 & & 14 &  & $1.15\pm0.10\phantom{^\dagger}$ \\
CRL 618 RN & 9.6 & & 6.65 & &      5 & $0.48\pm0.25^\dagger$ \\
DdDm 1 & 6.0 & & 31.3 & & 5 & \\
K 4-47 & 9.6 & & 4.29 & & 5 & $0.73\pm0.45^\dagger$\\
M 3-2 & 9.6 & & 2.86  & & 10 & $0.00\pm0.12^\dagger$\\
Me 2-1 & 12.0 & & 59.1 & & 5 & \\
PC 12 & 7.2 & & 188 & & 10  & \\
\end{tabular}

{\noindent \bf Table 1:}  Details of the observed sample of planetary
nebula.  The size of the region extracted along the slit is given in 
the column 2, the observed HI Br$\gamma$ fluxes from the separate
H and K band spectra in columns 3 and 4 (without any correction for
extinction), the derived 1$\sigma$ average error in the line fluxes
for the H$+$K and K band spectra in columns 5 and 6, the extinction 
at Br$\gamma$ in column 7.  We have separated
the objects into those where the object is compact or the spectrum
invariant along the slit, those with extended structure showing evidence
of different excitation conditions along the slit, and, finally, those 
objects only observed at K.  The regions labelled 
M 1-78 H$_2$~North and M 1-78 H$_2$~South
are directly to the north/south of the central H$^+$ nebula.
BD+30$^\circ$3639 Core is the region dominated by the WC central star.
The region labelled BD+30$^\circ$3639 Nebula is the H$^+$ 
region directly surrounding the core,
and BD+30$^\circ$3639 H$_2$ zone is the outer portion of the nebula
where H$^+$ emission is weak and H$_2$ strong.  The region labelled 
CRL 618 refers to the central obscured core of this nebula, whereas
CRL 618 RN is the brighter of the visible reflection nebula.

\tabcolsep=5pt
\begin{tabular}{llcccccccccc}
Line ($\mu$m) & ID  & Hu 1-2 & K 3-60 & K 3-62 & K 3-66 & K 3-67 & K 4-48 & M 1-1 & M 1-4 & M 1-6 & M 1-9\\ 
 1.61138 & HI     &  14.68  &  14.53  &  15.24  &  15.04  &  15.28  &  &  16.28  &  14.34  &  14.69 &  14.69\\ 
 1.64117 & HI/[FeII]   &  20.86  &  19.49  &  19.66  &  22.08  &  23.39  &  20.60  &  23.33  &  19.49  &  20.85&20.10\\ 
 1.68112 & HI     &  25.34  &  24.64  &  24.54  &  26.01  &  25.61  &  26.15  &  28.77  &  25.14  &  26.28&  25.49 \\ 
 1.69260 & HeII   &  13.70  &   7.55  &  &  &  &   3.52  &  15.63  &  & \\ 
 1.70071 & HeI    &   8.20  &   8.20  &  12.27  &  10.29  &  13.13  &  12.12  &  &  12.27  &   7.70 &11.99\\ 
 1.72680 & ?     &  &  &  &  &   1.29  &  &  &  & \\ 
 1.73669 & HI     &  34.57  &  35.30  &  33.51  &  32.44  &  32.63  &  33.09  &  39.74  &  34.92  &  33.17 &33.87\\ 
 1.74541/5 & [FeII]/HeI   &  &   1.61b  &   1.32  &   2.75  &   2.09  &  &  &   1.01  &   2.24 &1.04\\ 
 1.74795 & H$_2$     &  &   1.61b  &  &  &  &   8.45  &  &  & \\ 
 1.77170 & HeII   &   2.33  &  &  &  &  &  &  &  & \\ 
 1.78804 & H$_2$     &  &  &  &  &  &   6.36  &  &  & \\ 
 1.80909 & [FeII]/[FeIII]   &  &  &   1.66  &  &  &  &  &  & \\ 
 1.81791 & HI     &  70.54  &  57.88  &  89.22  &  59.48  &  51.44  &  85.50  &  80.21  & 105.11  &  86.17&34.12 \\ 
 1.94509 & HI     &  69.61  &  57.71  &  29.50  &  57.47  &  59.61  &  65.22  &  48.79  &  33.30  &  75.59&95.97 \\ 
 1.95485 & HeI    & p & p &   7.12  &  &   7.52  &  &  &   5.54  &   3.56 &8.06\\ 
 1.95756 & H$_2$     &  &  10.02  &  &  &  &  44.85  &  &  & \\ 
 1.96340 & [SiVI]  &  22.96  &  &  &  &  &  &  &  & \\ 
 2.03376 & H$_2$     &  &   4.70  &  &  &  &  17.34  &  &  & \\ 
 2.03790 & HeII   &   5.29  &   3.65  &  &  &  &  &  &  & \\ 
 2.0483 & ?      &  &  &  &  &  &  &  &  &   3.12 \\ 
 2.05870 & HeI    &  34.63  &  20.21  &  87.34  &  76.05  &  39.57  &  29.44  &  &  56.10  & 117.88 &95.79\\ 
 2.07347 & H$_2$     &  &  &  &  &  &   4.03  &  &  & \\ 
 2.113 & HeI (bl)   &   6.02  &   7.73  &   5.28  &   3.88  &   7.29  &   9.01  &  &   6.46  &   4.15 &5.33\\ 
 2.12183 & H$_2$     &   2.63  &  12.13  &  &  &   1.36  &  48.19  &  &  & \\ 
 2.15421 & H$_2$     &  &   1.69  &  &  &  &   4.63  &  &  & \\ 
 2.161 & HeI (bl)  &   3.29  &  &3.75  &   3.76  &   3.78  &  &  &   3.68  &   3.42 &3.98\\ 
 2.16613 & HI     & 100.00  & 100.00  & 100.00  & 100.00  & 100.00  & 100.00  & 100.00  & 100.00  & 100.00 &100.00 \\ 
 2.18910 & HeII   &  27.40  &  13.64  &  &  &  &   6.98  &  30.35  &   2.32  & \\ 
 2.20139 & H$_2$/UID  &  &   3.12  &   1.43  &  &  &   3.50  &  &  &   1.29 \\ 
\end{tabular}

\vspace*{0.5mm}

\begin{tabular}{llccccccccc}
Line ($\mu$m) & ID  & M 1-11 & M 1-12 & M 1-14 & M 1-20 & M 1-74 & NGC 7027 & PC 12 & SaSt 2-3 & Vy 1-1\\ 
 1.61138 & HI     &  14.86  &  14.81  &  14.73  &  14.89  &  14.53  &  13.83  &  13.81 & 16.13  &  15.44 \\ 
 1.64117 & HI/[FeII]     &  18.41  &  18.86  &  19.75  &  20.55  &  26.05  &  21.59  &  18.22&20.27  &  19.04 \\ 
 1.66698 & ?      &  &  &  &   1.04  &  &  &  & \\ 
 1.68112 & HI     &  25.27  &  25.45  &  26.02  &  25.50  &  25.95  &  26.05  &  24.98&24.59  &  24.22 \\ 
 1.69260 & HeII   &  &  &  &  &  &   7.81  &  & \\ 
 1.70071 & HeI    &  3.28  &   3.60  &  10.82  &  12.22  &  12.25  &  10.07  &  11.16&&  11.22 \\ 
 1.72880 & H$_2$     &   1.71  &   1.04  &   1.89  &  &  &   1.35  &  & \\ 
 1.73669 & HI     &  30.20  &  33.90  &  33.46  &  33.72  &  31.91  &  34.56  & 34.30& 32.72  &  33.97 \\ 
 1.74541/5 & [FeII]/HeI  &   3.00b  &   3.04  &   2.36  &   1.35  &   2.05b  &   1.82b  &  & \\ 
 1.74627 & H$_2$      &   3.00b  &    &    &     &2.05b    &   1.82b  &  & \\ 
 1.77170 & HeII   &  &  &  &  &  &   1.35  &  & \\ 
 1.80909 & H$_2$     &  &  &  &  &   3.78  &  &  & \\ 
 1.81791 & HI     &  87.79  &  47.41  &  71.73  &  31.57  &  96.59  &  91.16  &  90.39 &18.27  &  84.76 \\ 
 1.94509 & HI     &  88.30  &  98.01  &  76.91  &  88.57  &  83.47  &  82.85  &  24.85&77.02  &  36.04 \\ 
 1.95485 & HeI    &   5.72b  &   5.90  &   5.20  &  11.66  &  12.81b  &  11.20b  & 4.40&  &   5.71 \\ 
 1.95756 & H2     & 5.72b &  &  &  & 12.81b &  11.20b  &  &    \\ 
 2.03376 & H$_2$     & p &  &  &  & p &   4.08  &  & \\ 
 2.03790 & HeII   &  &  &  &  &  &   2.97  &  & \\ 
 2.049 & ?      &  &  &   2.75  &  &  &  &  & \\ 
 2.05870 & HeI    &  44.03  &  58.43  & 122.70  &  60.10  &  59.72  &  34.12  &  122.37 &26.61  &  38.79 \\ 
 2.07347 & H$_2$     & p &  &  &  &  &   1.21  &  & \\ 
 2.113 & HeI (bl)    &   1.95  &   1.68  &   4.47  &   6.22  &   5.18  &   7.35  & 4.21 & &   3.45 \\ 
 2.12183 & H$_2$     &   4.66  &  &  &  &   2.34  &  12.18  &  & \\ 
 2.15421 & H$_2$     &   3.73b  &     &  &     & 4.04b &   3.54b  &  &    \\ 
 2.161 & HeI (bl)    &   3.73b  &   2.88  &3.35  &   4.38  & 4.04b &   3.54b  & 5.42& &   3.52 \\ 
 2.16613 & HI     & 100.00  & 100.00  & 100.00  & 100.00  & 100.00  & 100.00  & 100.00 & 100.00  & 100.00 \\ 
 2.18910 & HeII   &  &  &  &  &  &  13.11  &  & \\ 
 2.20139 & H$_2$/UID     &   3.45  & p &  &  &  &3.85  &  & \\ 
\end{tabular}

\vspace*{-.8mm}

\noindent{\bf Table 2 (a):} Observed line fluxes after correction for
extinction from the H$+$K band data.  Lines for which the correct
identification is unknown are marked by a ?.  The lines marked as UID are those
discussed by Geballe, Burton \& Isaacman (1991).

{\tabcolsep=3pt
\begin{tabular}{llccccccc}
Line ($\mu$m) & ID  & CRL 618 & M 1-78 & M 1-78 & M 1-78 
& BD+30$^\circ$3639 & BD+30$^\circ$3639 & BD+30$^\circ$3639 \\ 
 & & & Nebula & H$_2$ North & H$_2$ South & Core & Nebula & H$_2$ zone \\
 1.59991 & [FeII]   &   6.26  &  &  &  &  &  & \\ 
 1.61138 & HI     &  15.33  &  14.52  &  &  &  15.17  &  17.10  & \\ 
 1.623 & CIII      &  &  &  &  &   1.67  &  & \\ 
 1.635/7 & CIII/CIV(bl)      &  &  &  &  &   1.65  &  & \\ 
 1.64117 & HI/[FeII]     &  59.56  &  29.28  & 316.17  &  45.87  &  18.99  &  26.64  &  46.84 \\ 
 1.663 & CIV      &  &  &  &  &   3.37  &  & \\ 
 1.67445 & ?      &  &  &  &  &   3.61  &  & \\ 
 1.67733 & [FeI]I   &  10.69  &  &  &  &  &  & \\ 
 1.68112 & HI     &  25.61  &  26.14  &  p&  34.33  &  24.76  &  28.99  & \\ 
 1.68777 & H$_2$      &  12.35  &  &  &  &  &  & \\ 
 1.69260 & HeII   &  &  &  &  &   5.58  & 2.19 & \\ 
 1.70071 & HeI    &   6.30  &  10.78  &  &  &  14.48  &   3.35  & \\ 
 1.71159 & [FeII]   &   4.93  &  &  &  &  &  & \\ 
 1.71474 & H$_2$      &   5.83  &  &  &  &  &  & \\ 
 1.72880 & H$_2$     &  &   0.91  &  &  &  &   2.27  & 64.71\\ 
 1.73593 & H$_2$     &  &   &  &  42.80b  &   &   & 104.31b \\ 
 1.73669 & HI     &  30.97  &  33.78  &  &  42.80b  &  45.35  &  35.99  & 104.31b \\ 
 1.74795 & H$_2$     &  25.52  &  4.04 & 247.29  &  60.37  &  &3.87  &52.68  \\ 
 1.785 & CII      &  &  &  &  &  38.66  & 3.07 & \\ 
 1.78804 & H$_2$     &  14.03  &   1.15  & 133.13  &  25.62  &  &  & \\ 
 1.800/1 & CIII/CIV   &  &  &  &  &  22.55  &  & \\ 
 1.80909 & H$_2$     &  &   4.23  &  &  &  &  & \\ 
 1.81791 & HI     &  67.85  &  82.74  & p & 114.57  &  97.77  &  87.99  &  96.48 \\ 
 1.94486 & H$_2$     & & & 159.08b  &  93.28b  &  &   &  94.25b \\ 
 1.94509 & HI     &  48.67  &  57.15  & 158.99b  &  93.26b  &  49.64  &  42.31  &  94.25b \\ 
 1.95485 & HeI    &  &  14.35b  &  &  &   6.10  &   5.39b  & \\ 
 1.95756 & H$_2$     &  65.49  &  14.32b  & 901.18  & 281.98  &  &   5.39b  & 155.21 \\ 
 2.03376 & H$_2$     &  30.57  &   3.76  & 249.34  &  80.27  &  &   2.40  &  65.76 \\ 
 2.05870 & HeI    &  72.22  &  89.76  &  92.77  &  92.31  &  32.69  &  19.04  &  34.64 \\ 
 2.06556 & H$_2$     &  &  &  &  &  &  &  38.87 \\ 
 2.0687 & CIV      &  &  &  &  &  14.26  &  & \\ 
 2.07347 & H$_2$     &  14.99  &   1.99  & 111.41  &  29.42  &  &   2.40  &  83.02 \\ 
 2.078 & CIV    &  &  &  &  &  32.33  &  & \\ 
 2.108/2.117 & CIII  &  &  &  &  &   45.07b  &  & \\ 
 2.113 & HeI (bl)   &   9.56  &   4.65  &  &  13.61  & 45.07b   &   2.30  & \\ 
 2.12183 & H$_2$     &  93.71  &  13.20  & 681.15  & 254.27  &  &   6.97  & 171.35 \\ 
 2.13671 & ?      &  10.20  &  &  &  &  &  & \\ 
 2.139 & CIV      &  &  &  &  &   2.29  &  & \\ 
 2.14570 & FeIII  &  19.65  &  &  &  &  &  & \\ 
 2.15421 & H$_2$     &   8.55b  &  & p & p &  &  &  42.37 \\ 
 2.15510 & FeIII  &   8.54b  &  &  &  &  &    &   \\ 
 2.161 & HeI (bl)   &  &   3.58  &  &  &   4.04  &   4.09  & \\ 
 2.16613 & HI     & 100.00  & 100.00  & 100.00  & 100.00  & 100.00  & 100.00  & 100.00 \\ 
 2.18207 & HeI    &  &  &  &  &   7.30  &  & \\ 
 2.18910 & HeII   &  &  &  &  &   3.90  &  & \\ 
\end{tabular}
}
\vspace*{3mm}

\noindent{\bf Table 2 (a) (ctd)}

\begin{tabular}{llccccccccc}
Line ($\mu$m) & ID  & Hu 1-2 & K 3-60 & K 3-62 & K 3-66 & K 3-67 & K 4-48 & M 1-1 & M 1-4 & M 1-6\\ 
 1.94509 & HI     &  53.05  &  51.00  &  28.72  &  42.40  &  50.26  &  73.26  &  42.08  &  32.40  &  98.61 \\ 
 1.95485 & HeI    &   5.33  &p  &   6.45  &  &   6.26  &  &  &   6.46  &   7.40 \\ 
 1.95756 & H$_2$     &  &  14.74  &  &  &  &  36.56  &  &  & \\ 
 1.96340 & [SiVI]  &  14.90  &  &  &  &  &  &  &  & \\ 
 2.03376 & H$_2$     &  &   4.01  &  &  &  &  19.89  &  &  & \\ 
 2.03790 & HeII   &   5.41  & p &  &  &  &  &  &  & \\ 
 2.0483 & ?      &  &  &  &  &  &  &  &  &   1.47 \\ 
 2.05870 & HeI    &  37.23  &  18.49  &  83.81  & 111.29  &  55.06  &  48.05  &  &  45.26  & 108.14 \\ 
 2.07347 & H$_2$     &  &  &  &  &  &   3.84  &  &  & \\ 
 2.1018 & ?      &  &  &  &  &  &  &  &  &   1.23 \\ 
 2.113 & HeI (bl)    &6.76  &6.71  &   5.89  &   3.32  &   7.58  &   7.39  &  &   6.99  &   3.69 \\ 
 2.12183 & H$_2$     &   3.85  &  12.82  &  &  &   1.87  &  52.30  &  &  & \\ 
 2.15421 & H$_2$     &  & p &  &  &  &   4.41  &  &  &    \\ 
 2.161 & HeI (bl)   &  &  &   2.70  &   2.11  &   2.73  &  &  &   2.81  & 2.19\\ 
 2.16613 & HI     & 100.00  & 100.00  & 100.00  & 100.00  & 100.00  & 100.00  & 100.00  & 100.00  & 100.00 \\ 
 2.18910 & HeII   &  25.13  &  13.85  &  &  &  &   6.47  &  32.07  &   2.49  & \\ 
 2.20139 & H$_2$/UID     &  &   3.77  &2.07  &  &  &   4.54  &  &   1.32  & 1.33\\ 
 2.21830 & FeIII  &  &  &  &  &   0.87  &  &  &  & \\ 
 2.22329 & H$_2$     &  &   3.40  &  &  &  &  13.20  &  &  & \\ 
 2.24772 & H$_2$     &  &  p&  &  &  &   5.85  &  &  & \\ 
 2.28705 & H$_2$/UID &  &   9.51  &   4.21  &  &   2.04  &  &  &   7.54  & \\ 
 2.34800 & HeII   &   8.69  &   4.99  &  &  &   &  &   9.55  &   1.05  & \\ 
 2.37438 & HI     &  &  &   0.98  &   2.02  &  &  &  &   1.28  &   0.89 \\ 
 2.38283 & HI     &  &  &   1.38  &   3.10  &   1.59  &  &  &   1.45  &   0.71 \\ 
 2.39247 & HI     &  &  &   1.74  &   1.59  &   1.69  &  &  &   1.71  &   1.81 \\ 
 2.40356 & HI     &  &  12.28b  &   1.97  &   3.47  &   2.63  &  &  &   2.24  &   1.34 \\ 
 2.40659 & H$_2$     &  &  12.28b  &  &  &     &  46.05b  &  &  & \\ 
 2.41434 & H$_2$     &  &   3.66b  &   &   &   &  17.07b  &  &   &  \\ 
 2.41639 & HI     &  &   3.66b  &   0.86  &   1.52b  &   1.51  &  17.06b  &  &   1.37  &   2.75 \\ 
 2.42373 & H$_2$     &  &   9.86  &  &  &   1.08  &  32.03  &  &   & \\ 
 2.43137 & HI     &  & p &   2.76  &   2.51  &   2.06  &  &  &   2.42  &   2.11 \\ 
 2.43749 & H$_2$     &  &  &  &  &  &  14.25  &  &  &    \\ 
 2.44901 & HI     &  & p &   4.28  &   5.30  &p  &   4.51  &  &   3.60  &   3.76 \\ 
 2.45475 & H$_2$     &  &  &  &  &  &  23.38  &  &  & \\ 
 2.46999 & HI     &  &p  &   4.50  &   4.52  & p &   8.20  &  &   2.91  &   4.02 \\ 
 2.47265 & ?      &  &  &  &  &  &   8.19  &  &  & \\ 
 2.47555 & H$_2$     &  & p &  &    &  &   4.71  &  &  & \\ 
\end{tabular}

\noindent{\bf Table 2 (b):} Observed line fluxes after correction for
extinction from the K band data.

\begin{tabular}{llccccccccc}
Line ($\mu$m) & ID  & M 1-9 & M 1-11 & M 1-12 & M 1-14 & M 1-20 & M 1-74 & NGC 7027 & SaSt 2-3 & Vy 1-1\\ 
 1.94509 & HI     &  84.98  &  75.45  & 102.74  &  97.99  &  81.51  &  84.89  &  70.47  &  89.03  &  27.78 \\ 
 1.95485 & HeI    &   4.35  & 4.29b &   2.18  &   5.99  &   8.92  &  17.93b  &   9.91b  &  & \\ 
 1.95756 & H$_2$     &  &   4.29b  &    &  &  &  17.92b  &   9.91b  &  & \\ 
 2.03376 & H$_2$     &  &   1.56  &  &  &  &   2.48  &   2.86  &  & \\ 
 2.03790 & HeII   &  &  &  &  &  &  &   3.31  &  & \\ 
 2.049 & ?      &2.11  &  &  &   1.40  &  &  &  &  & \\ 
 2.05870 & HeI    & 104.94  &  48.24  &  28.52  & 115.16  &  81.85  &  55.29  &  27.10  &  18.38  &  39.71 \\ 
 2.07347 & H$_2$     &  &   1.76  &  &  &  &  &   0.74  &  & \\ 
 2.102 & ?      &1.70  &1.02  &  &  &  &  &   0.68  &  & \\ 
 2.113 & HeI (bl)  &   5.83  &   1.30  &   1.54  &   4.39  &   5.95  &   5.19  & 7.24 &  & 5.45\\ 
 2.12183 & H$_2$     &  &   5.77  &    &  &  &   6.05  &   9.60  &  & \\ 
 2.138 & ?      &  &   0.78  &  &  &  &  &  &  & \\ 
 2.14570 & FeIII  &  &   0.96  &  &  &  &  &  &  & \\ 
 2.15421 & H$_2$     &  &   2.94b  &  &  &  &   2.70b  &   2.68b  &  & \\ 
 2.161 & HeI (bl)   &   2.94  & 2.94b &   2.07  &   2.08  &   2.30  & 2.70b & 2.68b &  &p \\ 
 2.16613 & HI     & 100.00  & 100.00  & 100.00  & 100.00  & 100.00  & 100.00  & 100.00  & 100.00  & 100.00 \\ 
 2.17939 & ?      &  &  &  &  &  &  &   1.13  &  & \\ 
 2.18910 & HeII   &  &  &  &  &  &  &  15.06  &  & \\ 
 2.20139 & H$_2$/UID     & p &   3.90  &   1.46  & p &  & p & 3.40 &  & \\ 
 2.21830 & FeIII  &  &  &  & p &  &   1.63  &   0.85  &  & \\ 
 2.22329 & H$_2$     &  &   2.06  &  &  &  &   2.34  &   2.80  &  & \\ 
 2.24270 & FeIII  &  &  &  &  &  &p  & p &  & \\ 
 2.24772 & H$_2$     &  &   2.11  &  &  &  &   2.13  & p &  & \\ 
 2.26095 & ?      &  &  &  &  &  &  &   0.78  &  & \\ 
 2.28705 & H$_2$/UID     &  &   0.97  &  &  &   2.11  &   3.23  &   8.39  &  & \\ 
 2.34800 & HeII   &  &  &  &  &  &   &   5.21  &  & \\ 
 2.34850 & FeIII  &  &  &  &  &  &   1.77  &  &  & \\ 
 2.35561 & H$_2$     &  & p &  &  &  &   1.38  & p &  & \\ 
 2.36694 & HI     & p &  &  &  &  & p &p  &  & \\ 
 2.37438 & HI     & p &  p  &  p  &  &   p  &   p  &  p  &  & \\ 
 2.38283 & HI     & p  & p  & p  &  & p &   p  &   p  &  & \\ 
 2.39247 & HI     &   2.82  &   1.97  &   1.72  &  &   1.46  &   1.74  &   1.40  &  & \\ 
 2.40356 & HI     &   3.31  &  6.73b  &   1.95  &  &   1.25  &   7.48b  &  12.10b  &  & \\ 
 2.40659 & H$_2$     &  &   6.73b  &    &  &  &   7.48b  &  12.10b  &  & \\ 
 2.41434 & H$_2$     &    &   6.12b  &   &  &  &   4.95b  &   5.01b  &  & \\ 
 2.41639 & HI     &   4.39  &   6.12b  &   3.76  &   1.45  &   2.59  &   4.95b  &   5.01b  &  & \\ 
 2.42373 & H$_2$     &  &   2.93  &  &  &  &   5.12  &   7.42  &  & \\ 
 2.43137 & HI     &   4.64  &   2.59  &   2.03  &   1.81  &   2.78  &   2.58  &   3.59  &  & \\ 
 2.43749 & H$_2$     &   &   4.14  &  &  &  &   2.40  &   1.85  &  & \\ 
 2.44901 & HI     &   4.09  &   5.89  &   4.64  &   2.49  &   4.09  &   5.10  &   4.08  &  & \\ 
 2.45280 & FeIII  &  &   2.40b  &  &  &  &   2.54b  & 1.80b &  & \\ 
 2.45475 & H$_2$     &  &   2.40b  &  &  &  &   2.54b  &   1.80b  &  & \\ 
 2.46125 & ?      &  &  &  &  &  &  &   1.58  &  & \\ 
 2.46999 & HI     &   5.79  &   4.93  &   4.46  &   3.16  &   3.57  &   5.65  &   5.31  &  & \\ 
\end{tabular}

\vspace*{3mm}

\noindent{\bf Table 2 (a) (ctd)} 

\begin{tabular}{llccccccccc}
Line ($\mu$m) & ID  & CRL 618 RN & DdDm 1 & K 4-47 & M 3-2 & Me 2-1 & PC 12\\ 
 1.94486 & H$_2$     & 208.86b  &   &  87.80b  & 262.25b  &   &  \\ 
 1.94509 & HI     & 208.83b  &  95.91  &  87.78b  & 262.25b  &  84.83  &  30.16 \\ 
 1.95485 & HeI    &  &   8.00  &  &  &  &   4.30 \\ 
 1.95756 & H$_2$     & 2940.81  &  & 1193.41  & 1012.58  &  & \\ 
 2.03376 & H$_2$     & 856.71  &  & 299.71  & 469.07  &  & \\ 
 2.03790 & HeII   &  &  &  &  &   4.42  & \\ 
 2.05870 & HeI    &  52.00  &  93.81  &  57.33  &  &   5.37  & 116.77 \\ 
 2.06556 & H$_2$     &  p  &  &  &  &  & \\ 
 2.07347 & H$_2$     & 239.48  &  &  67.46  & p &  & \\ 
 2.113 & HeI (bl)    &  40.32  & 5.75 &  &  & 5.30 &   3.35 \\ 
 2.12183 & H$_2$     & 2274.39  &  & 880.67  & 1286.96  & p &p \\ 
 2.15421 & H$_2$     & 102.34b  &  & p &  &  & \\ 
 2.15510 & FeIII  & 102.30b  &  &  &  &  & \\ 
 2.161 & HeI (bl)    &  &  &  &  &  &   2.36 \\ 
 2.16613 & HI     & 100.00  & 100.00  & 100.00  &  & 100.00  & 100.00 \\ 
 2.18910 & HeII   &  &  &  &  &  24.22  & \\ 
 2.20139 & H$_2$/UID     &  65.42  &  & p &  &  & \\ 
 2.21830 & FeIII  &  &   6.52  &  &  &  &p \\ 
 2.22329 & H$_2$     & 515.01  &  & 196.33  & 281.16  &  & \\ 
 2.24772 & H$_2$     & 241.90  &  &  73.90  & p &  & \\ 
 2.28705 & H$_2$/UID     & p   &  &p  & p &   4.38  & \\ 
 2.34454 & H$_2$     & p &  &  &  &  & \\ 
 2.34800 & HeII   &  &  &  &  &   7.12  &    \\ 
 2.34850 & FeIII  &  & p &  &  &   &   2.65 \\ 
 2.35561 & H$_2$     &  72.96  &  &  &  &  & \\ 
 2.37438 & HI     &  &  &  &  &  &   p \\ 
 2.38283 & HI     &   &  &  &  &  &   1.99 \\ 
 2.383 & ?     &  32.39  &  &  &  &  &    \\ 
 2.38649 & H$_2$     &  40.05  &  &  & p &  & \\ 
 2.39247 & HI     &  &  &  &  &  &   p \\ 
 2.39629 & ?      &  40.07  &  &  &  &  & \\ 
 2.40356 & HI     &  &  &  &  & p &   3.83 \\ 
 2.40659 & H$_2$     & 1725.77  &  & 717.84  & 911.92  &  &    \\ 
 2.41434 & H$_2$     & 608.19  &  & 245.30  & 337.82  &  & \\ 
 2.41639 & HI     &  &  &  &  & p &   3.08 \\ 
 2.42373 & H$_2$     & 1587.68  &  & 613.48  & 847.26  &  & \\ 
 2.43137 & HI     &  & p &  &  &   p  &   4.14 \\ 
 2.43749 & H$_2$     & 462.46  &  & 205.79  & 285.08  &  & \\ 
 2.446 & ?      &  &  &  & 145.72  &  & \\ 
 2.44901 & HI     &    &p  &  &  &p  &   5.29 \\ 
 2.45280 & FeIII  & 464.91b  &  & 195.68b  & 692.07b  &  & \\ 
 2.45475 & H$_2$     & 464.65b  &  & 195.51b  & 692.07b  &  & \\ 
 2.46999 & HI     &  &p  &  &   &p  &   6.78 \\ 
 2.47555 & H$_2$     & 291.56  &  & 106.65  &  &  & \\ 
\end{tabular}

\vspace*{3mm}

\noindent{\bf Table 2 (a) (ctd)} 

\begin{tabular}{llcccccc}
Line ($\mu$m) & ID  & M 1-78 & M 1-78 & M 1-78 
& BD+30$^\circ$3639 & BD+30$^\circ$3639 & BD+30$^\circ$3639 \\ 
 & & Nebula & H$_2$ North & H$_2$ South & Core & Nebula & H$_2$ zone \\
 1.94486 & H$_2$     &    &  92.18b  &  51.35b  &    &    &  46.91b \\ 
 1.94509 & HI     &  42.77  &  92.13b  &  51.34b  &  34.97  &  33.41  &  46.90b \\ 
 1.95485 & HeI    & 28.96b &  &  &   4.88  &   5.51b  & \\ 
 1.95756 & H$_2$     &  28.96b  & 351.30  & 108.18  &  &   5.51b  &  34.41 \\ 
 2.03376 & H$_2$     &   9.91  & 102.77  &  37.55  &  &   2.11  &  16.33 \\ 
 2.03790 & HeII   &  &  &  &   1.17  &  & \\ 
 2.04178 & H$_2$     &  &  &  &  &  &   5.36 \\ 
 2.05870 & HeI    &  79.80  &  87.54  &  78.49  &  32.38  &  23.43  &  27.27 \\ 
 2.06556 & H$_2$     &  &  &  &  &  &   p \\ 
 2.0687 & CIV      &  &  &  &   6.85  &  & \\ 
 2.07347 & H$_2$     &   4.07  &  44.38  &  13.61  &  &  &  16.35 \\ 
 2.078 & CIV      &  &  &  &   19.24  &  & \\ 
 2.113 & HeI (bl)    &   4.98  &   9.26  &   7.39  &  25.18  & 4.31 & \\ 
 2.12183 & H$_2$     &  36.14  & 311.45  & 125.16  &   1.45  &   6.19  &  42.40 \\ 
 2.139 & CIV      &  &  &  &   2.49  &  & \\ 
 2.15421 & H$_2$     &   &  15.89  &   7.76  &  &   2.80b  &  14.65 \\ 
 2.15510 & FeIII  &   &  15.86  &   7.75  & 2.88b   &   2.80b  &  14.65 \\ 
 2.161 & HeI    & 3.50 &  &  &   2.88b  & 2.80b  & \\ 
 2.16613 & HI     & 100.00  & 100.00  & 100.00  & 100.00  & 100.00  & 100.00 \\ 
 2.18207 & HeI    &  &  &  &   2.18  &  & \\ 
 2.18910 & HeII   &  &  &  &   3.04  &  & \\ 
 2.20139 & H$_2$/UID     &   1.83  &  10.51  &   4.33  &   0.90  &  &   9.23 \\ 
 2.21299 & ?      &  &   4.03  &  &  &  & \\ 
 2.21830 & FeIII  &   3.62  &  &   4.62  &   0.80  &  & \\ 
 2.22329 & H$_2$     &  10.74  &  70.65  &  28.34  &   1.85  &   1.77  &  12.77 \\ 
 2.24270 & FeIII  &   1.79  &  &  &  &  & \\ 
 2.24772 & H$_2$     &   4.19  &  29.35  &  10.65  &   1.04  & p &  17.03 \\ 
 2.26434 & ?      &  &  &  &   0.76  &  & \\ 
 2.278 & CIV      &  &  &  &   0.80  &  & \\ 
 2.28705 & H$_2$/UID     &   1.36  &   3.05  &  &  &  &   5.48 \\ 
 2.314 & HeII      &  &  &  &   0.54  &  & \\ 
 2.318/25 & CIII/CIV  &  &  &  &   8.94  &   1.20  & \\ 
 2.34454 & H$_2$     & p &   3.19  &  &    &  &   5.65 \\ 
 2.34800 & HeII   &   &    &  &   1.97  &  & \\ 
 2.34850 & FeIII  &   1.97  & p &  &    &  & \\ 
 2.35561 & H$_2$     &  &   9.69  &  &  &  &   9.26 \\ 
 2.36694 & HI     & p &  &  &   0.68  &  & \\ 
 2.37438 & HI     &p  &  &  &   1.78  &  & \\ 
 2.38283 & HI     &p  &    &  &   2.21  &   1.54  & \\ 
 2.38649 & H$_2$     &  &   7.88  &  &  &  &  10.72 \\ 
 2.39247 & HI     & p &    &  &   2.00  &   1.62  & \\ 
 2.39580 & ?      &  &   2.76  &  &  &  & \\ 
 2.40356 & HI     &    &  &  &   3.44b  &   6.58b  &   \\ 
 2.40659 & H$_2$     &  45.15  & 251.99  &  98.85  & 3.44b  & 6.58b  &  37.10 \\ 
 2.41434 & H$_2$     &  14.04b  &  85.69  &  32.02  &   1.81b  &   2.75b  &  19.46 \\ 
 2.41639 & HI     &  14.02b  &   &   &   1.81b  &   2.75b  &  \\ 
 2.42373 & H$_2$     &  31.04  & 210.43  &  82.95  &   2.60  &   4.24  &  28.97 \\ 
 2.43137 & HI     &   2.72  &  &  &   3.69  &   2.71  &   \\ 
 2.43749 & H$_2$     &   8.02  &  56.53  &  20.54  & p  &   1.93  &  13.06 \\ 
 2.44901 & HI     &   3.32  &  &  &   4.87  &   5.23  &  p \\ 
 2.45280 & FeIII  &  17.16b& 130.31b  &  48.93b  &   1.08b  &   2.02b  & pb \\ 
 2.45475 & H$_2$     &  17.16b  & 130.02b  &  48.89b  &   1.08b  &   2.01b  &   pb \\ 
 2.46999 & HI     &   4.30  &  &  &   6.66  &   6.54  &  p \\ 
 2.47555 & H$_2$     &   5.95  &  38.20  &  15.07  &  &  &   7.68 \\ 
\end{tabular}

\vspace*{3mm}

\noindent{\bf Table 2 (a) (ctd)}

\begin{tabular}{lcccc}
Name & \multicolumn{2}{c}{2.058$\mu$m HeI/Br$\gamma$ ratio}
	& \multicolumn{2}{c}{Correction Factor Applied}\\
 & H$+$K & K & H$+$K & K \\
Hu 1-2               & 0.3456 & 0.3665 & 1.0020 & 1.0157 \\
K 3-60               & 0.2413 & 0.2170 & 0.8113 & 0.8257 \\
K 3-62               & 0.7081 & 0.6976 & 1.1824 & 1.1516 \\
K 3-66               & 0.8730 & 0.9858 & 0.8600 & 1.1145 \\
K 3-67               & 0.4616 & 0.5045 & 0.8396 & 1.0689 \\
K 4-48               & 0.3495 & 0.4304 & 0.8137 & 1.0786 \\
M 1-1                & 0.0000 & 0.0000 & 0.0000 & 0.0000 \\
M 1-4                & 0.4492 & 0.4306 & 1.2028 & 1.0122 \\
M 1-6                & 1.0560 & 1.0080 & 1.0795 & 1.0374 \\
M 1-9                & 1.1134 & 1.0403 & 0.8494 & 0.9959 \\
M 1-11               & 0.4124 & 0.4088 & 1.0510 & 1.1614 \\
M 1-12               & 0.6176 & 0.3677 & 0.9340 & 0.7659 \\
M 1-14               & 1.0947 & 1.1063 & 1.1065 & 1.0277 \\
M 1-20               & 0.6051 & 0.7021 & 0.9825 & 1.1533 \\
M 1-74               & 0.6234 & 0.6011 & 0.9458 & 0.9080 \\
NGC 7027             & 0.3455 & 0.2850 & 0.9778 & 0.9418 \\
SaSt 2-3             & 0.2329 & 0.1572 & 1.1222 & 1.1483 \\
Vy 1-1               & 0.3148 & 0.3210 & 1.2102 & 1.2157 \\
M 1-78 Nebula        & 0.8266 & 0.7765 & 1.0114 & 0.9573 \\
M 1-78 H$_2$~North        & 0.7604 & 0.7581 & 1.0114 & 0.9573 \\
M 1-78 H$_2$~South        & 0.8501 & 0.7636 & 1.0114 & 0.9573 \\
BD+30$^\circ$3639 Core        & 0.3491 & 0.3394 & 0.9272 & 0.9446 \\
BD+30$^\circ$3639 Nebula      & 0.2033 & 0.2456 & 0.9272 & 0.9446 \\
BD+30$^\circ$3639 H$_2$ zone  & 0.3698 & 0.2888 & 0.9272 & 0.9446 \\
CRL 618              & 0.5747 &  & 1.1219 &  \\
CRL 618 RN           &  & 0.4878 &  & 1.0166 \\
DdDm 1               &  & 0.7686 &  & 1.2205 \\
PC 12                &  & 0.9182 &  & 1.2717 \\
\end{tabular}

\noindent{\bf Table 3:} 2.058$\mu$m 2$^1$P--2$^1$S HeI to HI Br$\gamma$
ratio after correction for extinction and atmospheric absorption as described
in the text.  The corrected values for the separate H$+$K and K band spectra
are given.  The actual correction factor applied is also given
for both wavebands.  The observed values are the product of the corrected
ratio and these values.  They are also given in Table 2 after
correction for extinction, but without correction for atmospheric absorption.

\begin{tabular}{lrr@{$\pm$}lr@{$\pm$}l}
Name               & \multicolumn{1}{c}{T$_{eff}$} & 
\multicolumn{2}{c}{I(HeI 2.058$\mu$m)/I(Br$\gamma$)} &
\multicolumn{2}{c}{I(HeI 1.701$\mu$m)/I(Br$\gamma$)} \\
BD303639 Nebula    &        25600& 0.227& 0.030& 0.034& 0.004\\
CRL 618            &        25600& 0.643& 0.006& 0.063& 0.006\\
DdDm 1             &        40000& 0.786& 0.016\\
Hu 1-2             &       300000& 0.356& 0.015& 0.082& 0.007\\
K 3-60             &       185000& 0.236& 0.018& 0.082& 0.004\\
K 3-62             &        57800& 0.733& 0.008& 0.123& 0.002\\
K 3-66             &        33100& 0.941& 0.081& 0.103& 0.003\\
K 3-67             &        59200& 0.493& 0.031& 0.131& 0.002\\
K 4-48             &       125000& 0.404& 0.059& 0.121& 0.007\\
M 1-1              &       340000& 0.000& 0.025& 0.000& 0.025\\
M 1-4              &        86000& 0.457& 0.014& 0.123& 0.002\\
M 1-6              &        29330& 1.067& 0.035& 0.077& 0.002\\
M 1-9              &        39900& 1.091& 0.052& 0.120& 0.003\\
M 1-11             &        25600& 0.417& 0.003& 0.033& 0.003\\
M 1-12             &        25890& 0.499& 0.179& 0.036& 0.003\\
M 1-14             &        34500& 1.115& 0.008& 0.108& 0.002\\
M 1-20             &        57000& 0.661& 0.069& 0.122& 0.002\\
M 1-74             &        60700& 0.620& 0.016& 0.122& 0.002\\
M 1-78 Nebula      &        35500& 0.860& 0.038& 0.108& 0.001\\
Me 2-1             &       280000& 0.054& 0.008\\
NGC 7027           &       190000& 0.318& 0.043& 0.101& 0.003\\
PC 12              &        35400& 0.966& 0.039& 0.127& 0.004\\
SaSt 2-3           &        28500& 0.199& 0.055& 0.000& 0.013\\
Vy 1-1             &        56200& 0.324& 0.004& 0.112& 0.006\\
\end{tabular}

\noindent{\bf Table 4:} Final adopted values of the effective temperature,
2.058$\mu$m 2$^1$P--2$^1$S HeI to HI Br$\gamma$ ratio and 1.7007$\mu$m
4$^{3}$D--3$^{3}$P HeI to HI Br$\gamma$ ratio for our dataset.  Results are
shown for all PN plotted in Figure 4.

\begin{tabular}{lcc}
Name & T$_{ex} (K) $ & O/P\\

\multicolumn{3}{c}{ v=1--0 transitions}\\

BD+30$^\circ$3639 Nebula      & 1240$\pm$370 & 2.64$\pm$0.69\\
BD+30$^\circ$3639 H$_2$ zone  & 1440$\pm$200 & 2.44$\pm0.21$\\
CRL 618              & 1390$\pm60$ \\
CRL 618 RN           & 1890$\pm50$  & 2.87$\pm0.04$\\
K 3-60               & 1560$\pm420$ & 2.85$\pm0.99$\\
K 4-47               & 1550$\pm140$ & 3.02$\pm0.03$\\
K 4-48               & 1720$\pm120$ & 2.68$\pm0.14$\\
M 1-11               & 760$\pm50$   & 2.55$\pm0.21$\\
M 1-74               & 1140$\pm210$ & 2.02$\pm0.34$\\
M 1-78 Nebula        & 980$\pm25$   & 2.82$\pm0.11$\\
M 1-78 H$_2$~North        & 1530$\pm$70  & 3.02$\pm0.05$\\
M 1-78 H$_2$~South        & 1310$\pm$60  & 3.17$\pm$0.10\\
M 3-2                & 1810$\pm360$ & 2.97$\pm$0.35\\
NGC 7027             & 1090$\pm$90  & 2.74$\pm0.25$\\

\multicolumn{3}{c}{ v=2--1 transitions}\\

BD+30$^\circ$3639 H$_2$ zone  & 1300$\pm$110 & 1.24$\pm0.12$\\
                     &              & 1.15$\pm0.14$\\
CRL 618 RN           & 1800$\pm100$ & 2.62$\pm0.23$\\
                     &              & 2.32$\pm0.37$\\
K 4-47               & 1720$\pm540$ & 3.5$\pm0.3$ \\
K 4-48               & 1290$\pm170$\\
M 1-11               & 1370$\pm130$\\
M 1-78 Nebula        & 2400$\pm710$\\
M 1-78 H$_2$~North        & 3580$\pm$710 & 2.47$\pm0.21$ \\
                     &              & 1.98$\pm$0.23 \\
M 1-78 H$_2$~South        & 3300$\pm$670 & 1.71$\pm0.23$ \\

\multicolumn{3}{c}{ v=3--2 transitions}\\

BD+30$^\circ$3639 H$_2$ zone  & 3500$\pm$900\\
CRL 618 RN           & 4200$\pm1400$\\
M 1-78 H$_2$~North        & 3900$\pm$3100\\
\end{tabular}

\noindent{\bf Table 5:} Derived rotational excitation temperatures
and ortho/para ratios for the molecular hydrogen emission
from the objects shown in Figure 6.  The excitation
temperatures are derived from the fits to the slope of the individual
sets of transitions in the right hand panels of Figure 6.  The ortho/para
ratios are derived according to the prescription of Ramsay et al.\ (1993).
For the v=2--1 transitions, two ortho/para ratios were derived.
The first is from the comparison of the
2--1 S(3) and 2--1 S(2) lines, the second from the 2--1 S(1) and 2--1 S(2)
lines.  Where only one value is given it was derived from the 2--1 S(3) and
2--1 S(2) lines.

\newpage

\begin{center}
\begin{minipage}{7in}{
\psfig{file=Figure1a.ps,width=6.5in,angle=0,clip=}
}\end{minipage}
\vspace*{-2mm}
\end{center}

\vspace*{-10mm}

\begin{center}
\begin{minipage}{7in}{
\psfig{file=Figure1b.ps,width=6.5in,angle=0,clip=}
}\end{minipage}
\vspace*{-2mm}
\end{center}

{\noindent \bf Figure 1(a):} H band spectra of our targets.  The first
two panels show those objects without molecular hydrogen emission.

\begin{center}
\begin{minipage}{7in}{
\psfig{file=Figure1c.ps,width=6.5in,angle=0,clip=}
}\end{minipage}
\vspace*{-2mm}
\end{center}

\vspace*{-10mm}

\begin{center}
\begin{minipage}{7in}{
\psfig{file=Figure1d.ps,width=6.5in,angle=0,clip=}
}\end{minipage}
\vspace*{-2mm}
\end{center}

{\noindent \bf Figure 1(a) (ctd):} H band spectra of our targets.  The last two
panels show those objects which show significant molecular hydrogen emission at
K (see Table 5), with the exception of SaSt~2-3.  The data are truncated
for the plots of the molecular hydrogen emission regions of M~1-78 and
BD+30$^\circ$3639 as the data shortwards of 1.63$\mu$m are very noisy.
We have averaged the two molecular hydrogen emission regions for M~1-78 
to enhance the quality.

\begin{center}
\begin{minipage}{7in}{
\psfig{file=Figure1e.ps,width=6.5in,angle=0,clip=}
}\end{minipage}
\vspace*{-2mm}
\end{center}

\vspace*{-10mm}

\begin{center}
\begin{minipage}{7in}{
\psfig{file=Figure1f.ps,width=6.5in,angle=0,clip=}
}\end{minipage}
\vspace*{-2mm}
\end{center}

{\noindent \bf Figure 1(b):} K band spectra of our targets.  The first
two panels show those objects without molecular hydrogen emission.

\begin{center}
\begin{minipage}{7in}{
\psfig{file=Figure1g.ps,width=6.5in,angle=0,clip=}
}\end{minipage}
\vspace*{-2mm}
\end{center}

\vspace*{-10mm}

\begin{center}
\begin{minipage}{7in}{
\psfig{file=Figure1h.ps,width=6.5in,angle=0,clip=}
}\end{minipage}
\vspace*{-2mm}
\end{center}

{\noindent \bf Figure 1(b) (ctd):} K band spectra of our targets.  
The last two panels show those objects which show significant 
molecular hydrogen emission at
K (see Table 5), with the exception of SaSt~2-3.  
We have averaged the two molecular hydrogen emission regions for M~1-78.

\begin{center}
\begin{minipage}{7in}{
\psfig{file=Figure1i.ps,width=6.5in,angle=0,clip=}
}\end{minipage}
\vspace*{-2mm}
\end{center}

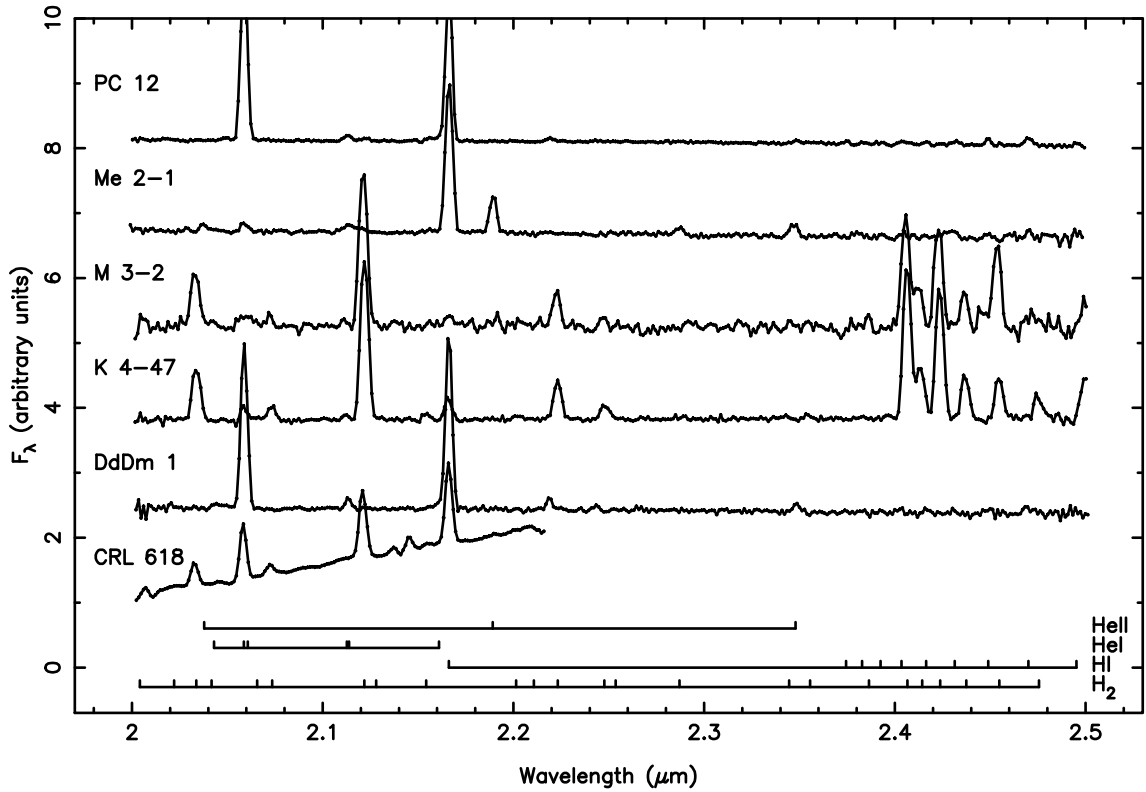
{\noindent \bf Figure 1(c):} H$+$K and K band spectra for those PN
only observed at one wavelength.  The K band spectra of CRL~618
is truncated since the H$+$K band data only cover the region
out to 2.2$\mu$m.  The other objects were only observed at K.

\newpage

\begin{center}
\begin{minipage}{7in}{
\psfig{file=Figure2.ps,width=6in,angle=0,clip=}
}\end{minipage}
\vspace*{-2mm}
\end{center}

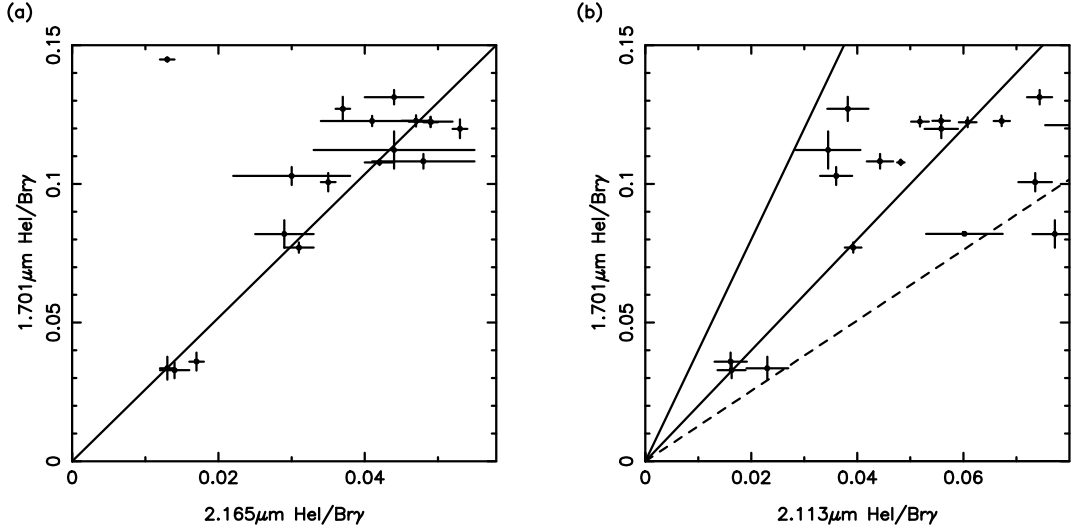
\noindent{\bf Figure 2:} (a) The observed 2.16475$\mu$m $7^{3,1}$G--$4^{3,1}$F
HeI line plotted against 1.7007$\mu$m 4$^{3}$D--3$^{3}$P HeI line, after
correction for extinction and ratioing both with HI Br$\gamma$.  The solid line
is the expected theoretical trend for these lines.  The highly discrepant point
arises in the central Wolf-Rayet spectrum of BD+30$^\circ$3639.  (b) The
observed 2.113$\mu$m 4$^{3,1}$S--3$^{3,1}$P HeI blend plotted against
1.7007$\mu$m 4$^{3}$D--3$^{3}$P HeI line in the same fashion as in (a).  Here
the solid lines represent the range of theoretical values allowed in a low
density nebula (one in which collisions are unimportant).  The dashed line
represents the value for higher density nebula with $n_e=10^4$cm$^{-3}$ and
shows the effect collisional excitation can have on the 2.113$\mu$m line.

\begin{center}
\begin{minipage}{7in}{
\psfig{file=Figure3.ps,width=6in,angle=0,clip=}
}\end{minipage}
\vspace*{-2mm}
\end{center}

\vspace*{-2mm}

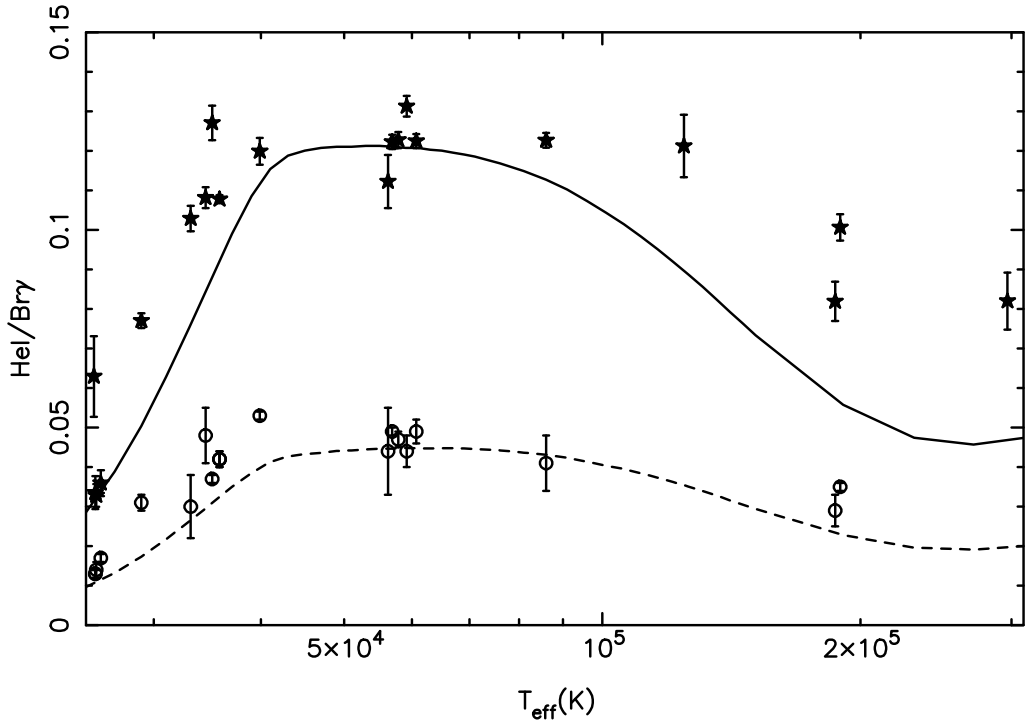
\noindent{\bf Figure 3:} Ratios of the 2.16475$\mu$m HeI
$7^{3,1}$G--$4^{3,1}$F line (open circles) and 
1.7007$\mu$m 4$^{3}$D--3$^{3}$P HeI line (asterisks) with Br$\gamma$
for the PN in our sample.  The stellar temperatures are taken from
Table 1 in Paper 1.  The solid line is the predicted ratio from a
suite of Cloudy models for the 1.7007$\mu$m line.  The dashed line
is from the same models but for the 2.16475$\mu$m line.  Note the clear
deviations between observation and model for the 1.7007$\mu$m line
at both low and high effective temperature.  The same trends may
also be evident in the somewhat lower signal-to-noise data for the
2.16475$\mu$m line.

\vspace*{-15mm}

\begin{center}
\begin{minipage}{7in}{
\psfig{file=Figure4.ps,width=6in,angle=0,clip=}
}\end{minipage}
\vspace*{-2mm}
\end{center}

\vspace*{-2mm}

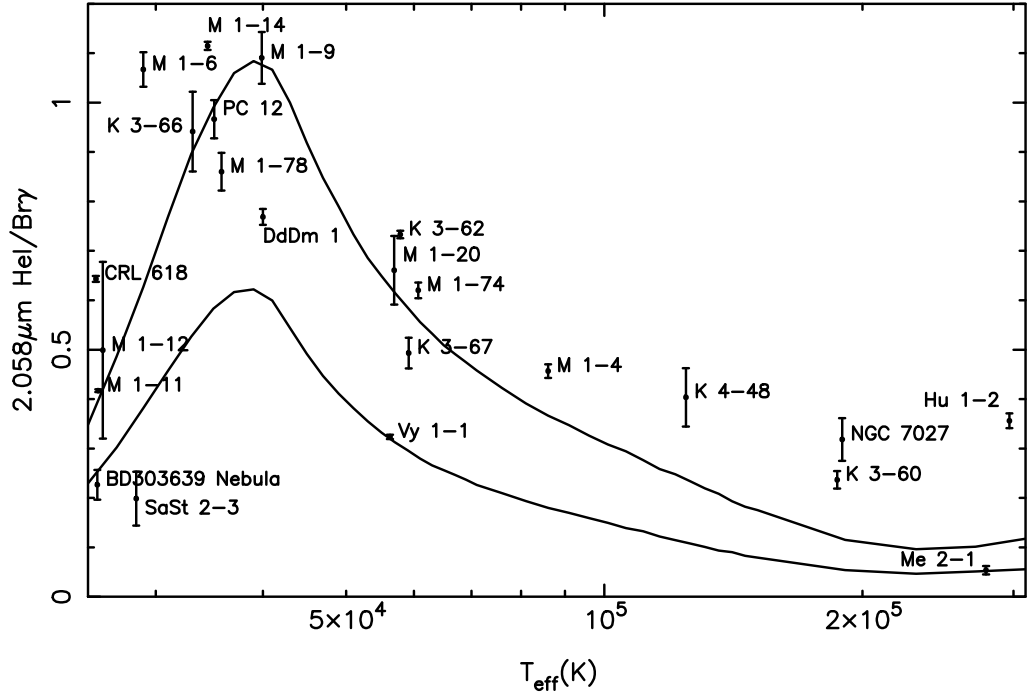
{\bf Figure 4:} The observed and predicted behaviour of the ratio of the
2.058$\mu$m HeI $2^{1}$P--$2^{1}$S line with HI Br$\gamma$.  The observed data
have been corrected for extinction.  The models plotted are for
$n_e=48000$cm$^{-3}$ and v$_{turb}=0$kms$^{-1}$ (upper curve) and
$n_e=3000$cm$^{-3}$ and v$_{turb}=15$kms$^{-1}$ (lower curve) which are
essentially the extreme maxima and minima in our model grids, as described in
Paper 1.  In both cases L$_*=5000$\Lsolar.  T$_{eff}$ is taken from Table 1 in
Paper 1.  The predictions from Cloudy are clearly not a good match to
the observed data.

\vspace*{-10mm}

\begin{center}
\begin{minipage}{7in}{
\psfig{file=Figure5.ps,width=6in,angle=0,clip=}
}\end{minipage}
\vspace*{-2mm}
\end{center}

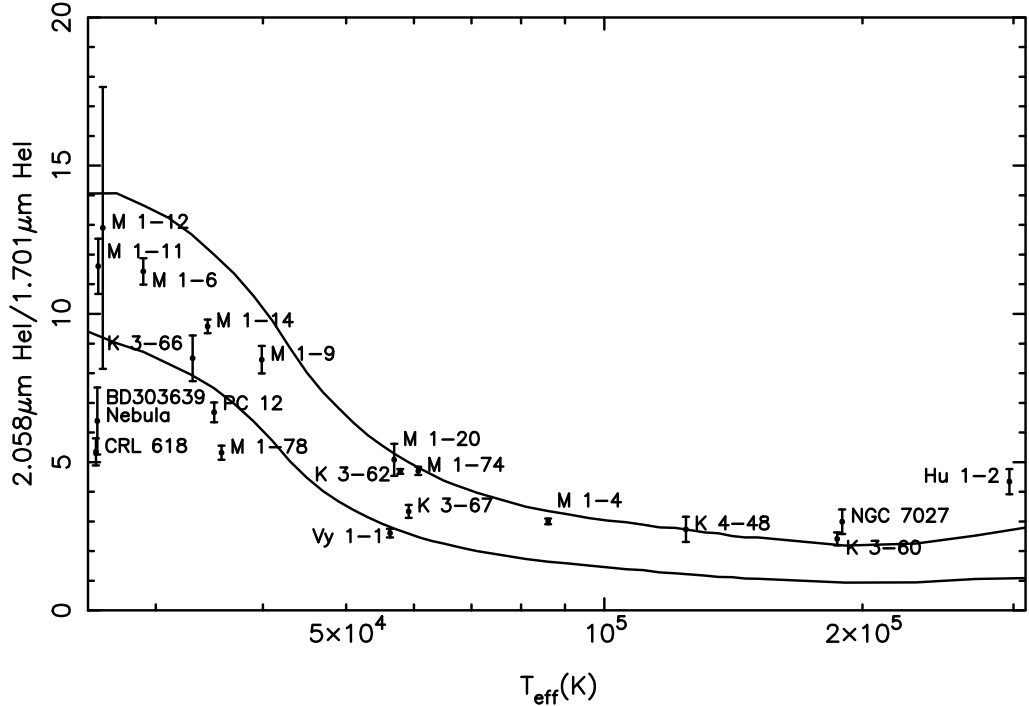
{\bf Figure 5:} The observed and predicted behaviour of the ratio of the
2.058$\mu$m HeI $2^{1}$P--$2^{1}$S line with the 1.7007$\mu$m
4$^{3}$D--3$^{3}$P HeI line.  The models plotted are the same as in Figure 4.
Here the observations do agree rather better with the models, which may
indicate that it is the He$^+$ fraction that is discrepant in our models.

\begin{center}
\begin{minipage}{6.9in}{\hspace*{3mm}(a)}\end{minipage}
\vspace*{-16mm}

\begin{minipage}{7in}{
\psfig{file=Figure6a.ps,width=6in,angle=0,clip=}
}\end{minipage}
\vspace*{2mm}
\end{center}

\begin{center}
\begin{minipage}{6.9in}{\hspace*{3mm}(b)}\end{minipage}
\vspace*{-16mm}

\begin{minipage}{7in}{
\psfig{file=Figure6b.ps,width=6in,angle=0,clip=}
}\end{minipage}
\vspace*{2mm}
\end{center}

\begin{center}
\begin{minipage}{6.9in}{\hspace*{3mm}(c)}\end{minipage}
\vspace*{-16mm}

\begin{minipage}{7in}{
\psfig{file=Figure6c.ps,width=6in,angle=0,clip=}
}\end{minipage}
\vspace*{2mm}
\end{center}

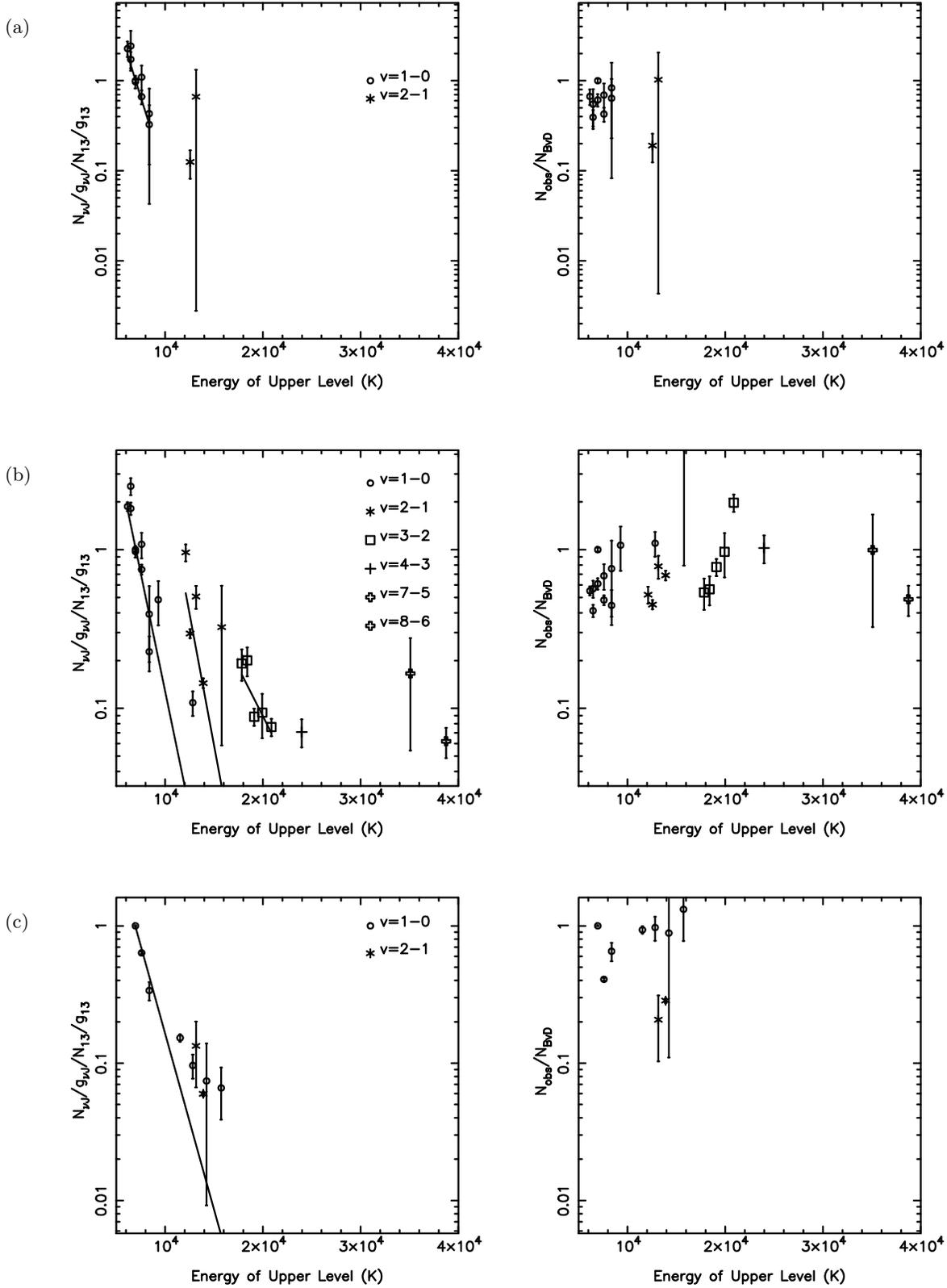
{\bf Figure 6:} The observed molecular hydrogen column densities 
divided by their relevant statistical weights, and scaled to
the observed 1--0 S(1) data are shown in the left hand panels.
The separate vibrational sequences are indicated.  
The same data are shown in the right hand panels but after
division by the predictions of model 14 of Black \& van Dishoeck (1987).
The plots are for: (a) the BD+30$^\circ$3639 nebula; (b)
the BD+30$^\circ$3639 H$_2$ emission region; (c) the central CRL~618
H band data.

\begin{center}
\begin{minipage}{6.9in}{\hspace*{3mm}(d)}\end{minipage}
\vspace*{-16mm}

\begin{minipage}{7in}{
\psfig{file=Figure6d.ps,width=6in,angle=0,clip=}
}\end{minipage}
\vspace*{2mm}
\end{center}

\begin{center}
\begin{minipage}{6.9in}{\hspace*{3mm}(e)}\end{minipage}
\vspace*{-16mm}

\begin{minipage}{7in}{
\psfig{file=Figure6e.ps,width=6in,angle=0,clip=}
}\end{minipage}
\vspace*{2mm}
\end{center}

\begin{center}
\begin{minipage}{6.9in}{\hspace*{3mm}(f)}\end{minipage}
\vspace*{-16mm}

\begin{minipage}{7in}{
\psfig{file=Figure6f.ps,width=6in,angle=0,clip=}
}\end{minipage}
\vspace*{2mm}
\end{center}

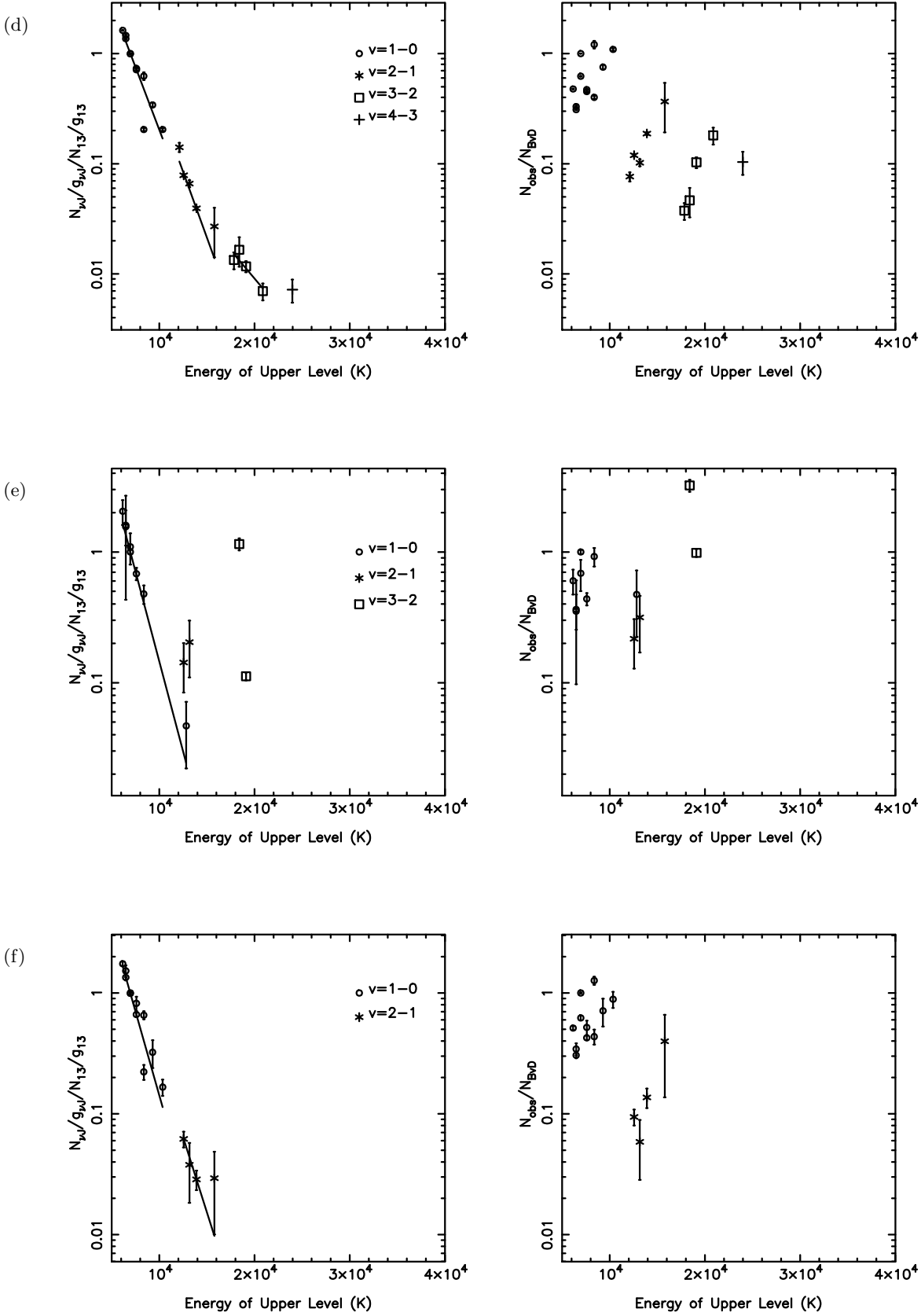
\noindent{\bf Figure 6 (ctd):} 
The plots are for: (d) the CRL~618 reflection nebula; (e) K~3-60;
(f) K~4-47.

\begin{center}
\begin{minipage}{6.9in}{\hspace*{3mm}(g)}\end{minipage}
\vspace*{-16mm}

\begin{minipage}{7in}{
\psfig{file=Figure6g.ps,width=6in,angle=0,clip=}
}\end{minipage}
\vspace*{2mm}
\end{center}

\begin{center}
\begin{minipage}{6.9in}{\hspace*{3mm}(h)}\end{minipage}
\vspace*{-16mm}

\begin{minipage}{7in}{
\psfig{file=Figure6h.ps,width=6in,angle=0,clip=}
}\end{minipage}
\vspace*{2mm}
\end{center}

\begin{center}
\begin{minipage}{6.9in}{\hspace*{3mm}(i)}\end{minipage}
\vspace*{-16mm}

\begin{minipage}{7in}{
\psfig{file=Figure6i.ps,width=6in,angle=0,clip=}
}\end{minipage}
\vspace*{2mm}
\end{center}

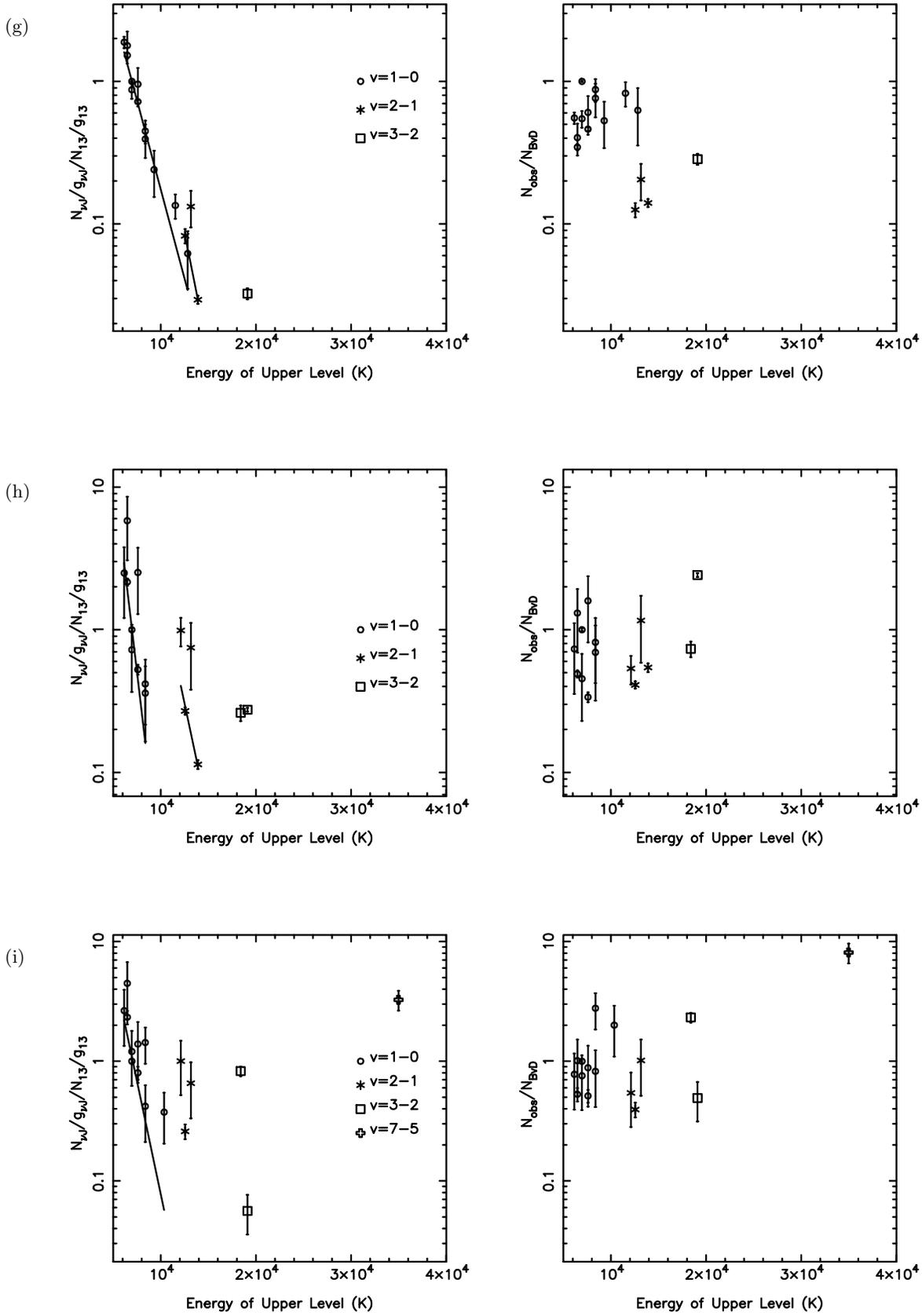
\noindent{\bf Figure 6 (ctd):} 
The plots are for: (g) K~4-48; (h) M~1-11; (i) M~1-74.

\begin{center}
\begin{minipage}{6.9in}{\hspace*{3mm}(j)}\end{minipage}
\vspace*{-16mm}

\begin{minipage}{7in}{
\psfig{file=Figure6j.ps,width=6in,angle=0,clip=}
}\end{minipage}
\vspace*{2mm}
\end{center}

\begin{center}
\begin{minipage}{6.9in}{\hspace*{3mm}(k)}\end{minipage}
\vspace*{-16mm}

\begin{minipage}{7in}{
\psfig{file=Figure6k.ps,width=6in,angle=0,clip=}
}\end{minipage}
\vspace*{2mm}
\end{center}

\begin{center}
\begin{minipage}{6.9in}{\hspace*{3mm}(l)}\end{minipage}
\vspace*{-16mm}

\begin{minipage}{7in}{
\psfig{file=Figure6l.ps,width=6in,angle=0,clip=}
}\end{minipage}
\vspace*{2mm}
\end{center}

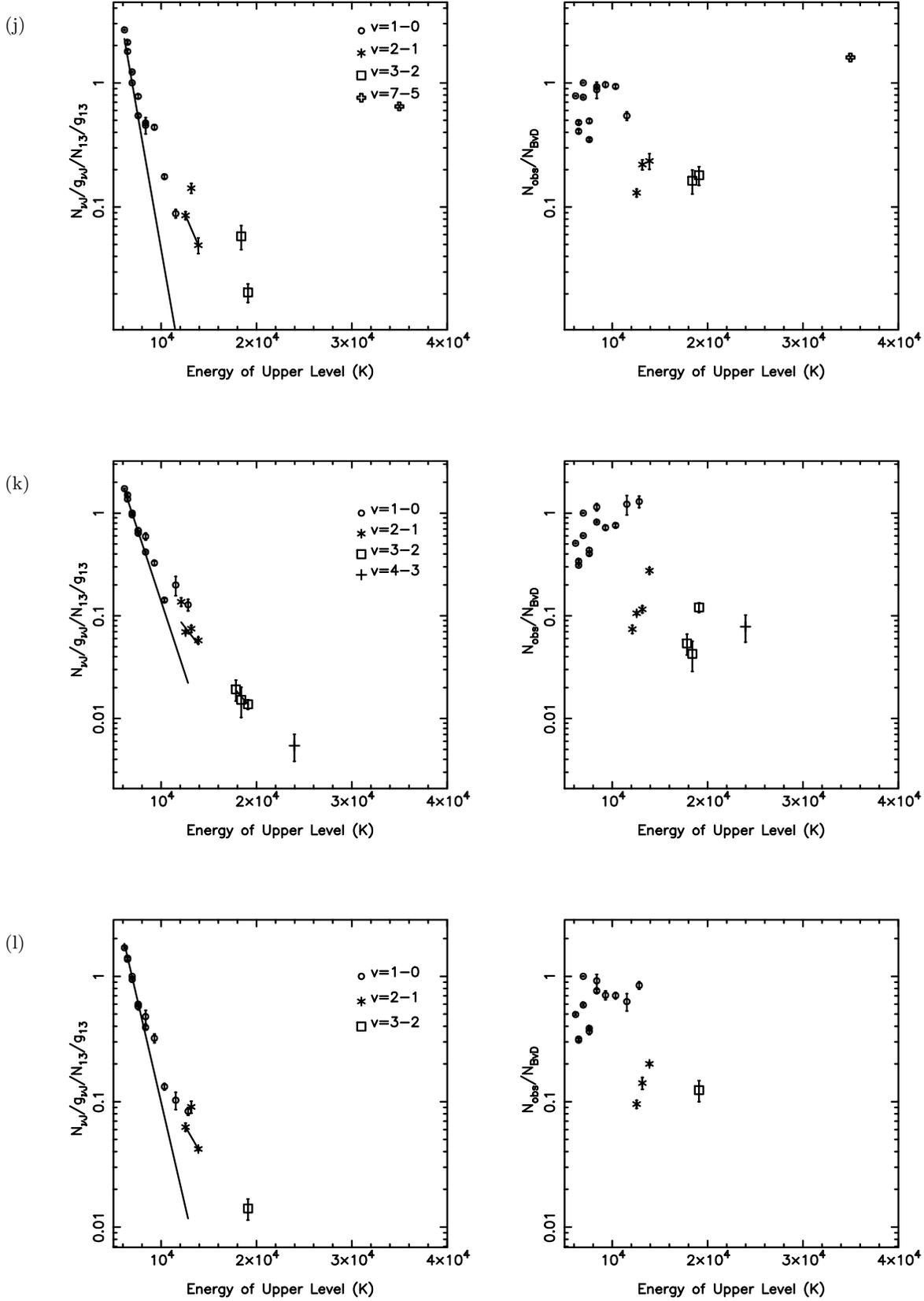
\noindent{\bf Figure 6 (ctd):} 
The plots are for: (j) the M~1-78 nebula; (k) the northern M~1-78
H$_2$ emission zone; (l) the southern M~1-78
H$_2$ emission zone.

\begin{center}
\begin{minipage}{6.9in}{\hspace*{3mm}(m)}\end{minipage}
\vspace*{-16mm}

\begin{minipage}{7in}{
\psfig{file=Figure6m.ps,width=6in,angle=0,clip=}
}\end{minipage}
\vspace*{2mm}
\end{center}

\begin{center}
\begin{minipage}{6.9in}{\hspace*{3mm}(n)}\end{minipage}
\vspace*{-16mm}

\begin{minipage}{7in}{
\psfig{file=Figure6n.ps,width=6in,angle=0,clip=}
}\end{minipage}
\vspace*{2mm}
\end{center}

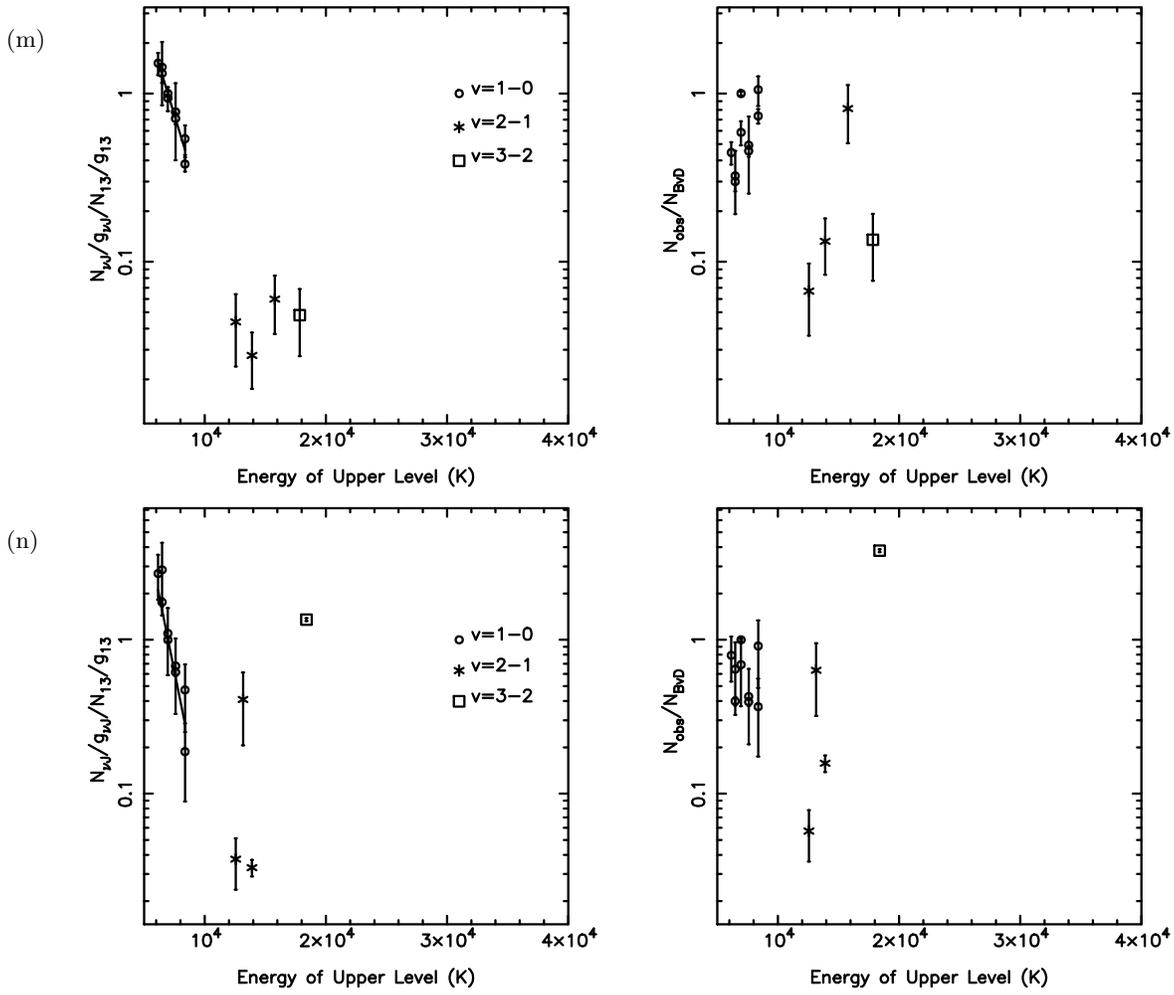
\noindent{\bf Figure 6 (ctd):} 
The plots are for: (m) M~3-2; (n) NGC~7027.

\end{document}